\documentclass[a4paper,11pt]{article}
\usepackage{jcappub} 
\usepackage{txfonts}
\usepackage{mathtools}

\title{\boldmath Observational tests in scale invariance I: galaxy  clusters and  rotation of  galaxies}

\author[]{Andre Maeder }

\affiliation[]{Department  of Astronomy, University of Geneva - chemin des Maillettes, 51, CH-1290 Sauverny, Switzerland}

\emailAdd{andre.maeder@unige.ch}

\abstract{
Galaxy velocities in clusters, rotation curves of galaxies, and "vertical" oscillations in the Milky Way itself currently show too high velocities with respect to the masses thought to be involved. While these velocity excesses are currently interpreted as the consequence of dark matter, it can also be naturally explained as a consequence of scale invariant theory, which rests on a very simple first principle: the addition of a new fundamental symmetry. The case of scale invariance, present in General Relativity and Maxwell equations for the empty space without charge and current, is considered. Cosmological models predict a rapid decrease of these effects with increasing mean density up to the critical density, where they totally disappear. Starting from the scale invariant geodesic equation by Dirac (1973), for which a demonstration by an action principle is presented, a modified Newton equation is derived. 
The solutions of  this equation are applied to clusters of galaxies, galactic rotation and  "vertical" motions in the Milky Way. In this new framework the following results emerge:
 - 1. The mass estimates of galaxy clusters are very  much  reduced with respect to the standard case.
 - 2. The {\bfseries{relatively}} flat rotation curves of galaxies {\bfseries{is favoured by}} the dynamical evolution.  
 - 3. Steeper rotation curves are predicted in significantly redshifted galaxies,  in a consistent way with recent observations showing less dark matter at higher $z$. 
 - 4. In clusters of galaxies and in galaxies, both at low and  high $z$, the amount of dark matter necessary to account for the velocities seems to be always determined by the amount of baryonic matter,  which points in favour of a gravity effect.
- 5. The growth of  the "vertical"  velocity dispersion, i.e., perpendicular to the plane, with stellar  ages  in the Milky Way is also positively tested.
The  convergence of  these results,  in very different gravitational systems, epochs, mass ranges and spatial scales,  opens very interesting perspectives that deserve to be explored further.}

\begin{document}
\maketitle
\flushbottom

\section{Introduction}

It is a general trend in dynamical systems that the observed velocities are  too high with respect to current  mass estimates. 
This was first observed in clusters of galaxies by Zwicky \cite{Zwicky33},   and later by Karachantsev \cite{Karachantsev66} 
who obtained mass excesses reaching a factor 80. Further
studies have well confirmed the above trend, for example mass-luminosity ratios up
to  a  factor 700 were found by Bahcall \cite{Bahcall74},  while Newtonian Virial masses reaching factors 30-50 were
estimated by Proctor \cite{Proctor15} for large clusters.  Noticeably,  the total dynamical
 masses of clusters  (and galaxies) appears, as shown by Chan  \cite{Chan22}, 
 always related to  the baryonic masses of the central regions of the clusters.

The rotation curves of galaxies raised several  problems since the early discovery 
of flat rotation curves in spiral galaxies up to galactocentric distances of a few tens of kpcs
 \cite{Rubin78}. A complete review of the observations  and their 
analysis has been given by Sofue and Rubin \cite{Sofue01}, while the challenges 
to the dark matter interpretation 
have also been reviewed  \cite{Famaey12}.
 The flat rotation curves played a major role in the raise of the hypothesis of dark matter (DM), which, 
with the accumulation of observations,  left the characteristics
of an astronomical hypothesis  of the late 70's to  progressively get in the 21$^{st}$ century the status  of a material  reality. {\bfseries{Recently new determinations of the rotation curves from 6D  (locations and velocities)
analysis by Eilers et al. \cite{Eilers19} and  by Jiao et al. \cite{Jiao23} have shown   significant  declines of the velocities between 5 and 25 kpc from the galactic center.}}

The radial acceleration relation  (RAR) studied by Lelli et al. \cite{Lelli17} compares
the  centrifugal accelerations $\frac{\upsilon^2}{R}$   and  the gravity 
$\frac{GM}{R^2}$ (masses being estimated from the luminosities)
in different locations of a large number of galaxies of different types. The corresponding points  are deviating from a 1:1 relation  
for gravities below a gravity  $a_0= 1.2 \times 10^{-8}$  cm s$^{-2}$. Such properties,  in particular the flat rotation curves 
were at the origin of the MOND theory by Milgrom, suggesting
 a modification of the gravity law \citep{Milgrom83,Milgrom09}. Several observations in favour of MOND have also been
identified \cite{Famaey12}. The deviations
from the  1:1 relation are taken as an indication of the amount of dark matter.
 A most puzzling fact is that the amount  of dark matter appears to be always
  directly determined by the amount of baryonic matter \cite{Lelli17}, although a total of 2700 measurements
  have been made in  galaxies of different types and at different distances from their centers. Such 
a strong connection between DM and baryons appears as a general property, it was also
recently found in clusters  \cite{Chan22} and in a large sample of massive star forming galaxies
at $z=0.6-2.5$ by Nestor Shachar et al. \cite{Nestor Shachar23}.

The recent observations of rotation curves in galaxies at the peak of their star formation
is bringing new questions. Surveys by Genzel et al. \cite{Genzel17} and Lang et al. \cite{Lang17} 
showed that the  rotation  curves  in galaxies with redshifts $z$ between 0.6 and 2.6
were not flat, but steeply decreasing with radius, a point which  raised much discussions 
before being well confirmed by a further analysis \cite{Genzel20}. Recently new observations
with ESO-VLT, LUCI, NOEMA at IRAM and ALMA of hundreds of galaxies at redshifts 
$z=0.6-2.5$ have confirmed the steep decrease of the velocity curves with  galactocentric distance in the past,
 thus implying according to the authors a decrease of the amount of dark matter
down to 17\% at $z=2.44$  \cite{Nestor Shachar23}. These results, based on new observations of
a larger sample of galaxies, give a remarkable support to the earlier findings by the Genzel  group. 
They raise new questions, since this picture does not appear to correspond to the $\Lambda$CDM models,
where the baryonic matter sets into the potential well early formed  by DM.  

Within galaxies,  several  critical  dynamical problems were already found long ago. Spitzer and Schwarzschild 
  \cite{Spitzer51} showed that the  "vertical" velocity dispersion (perpendicular to the
galactic plane) was much increasing with the age of the stellar group considered and this was interpreted
as a due to collisions with large interstellar clouds.
 This  dispersion increase  was  confirmed by  further  studies \cite{Seabroke07}. Since 75 years, many other
astrophysical effects were  proposed to account for the growth of velocities with age.

The  aim  of this work is to examine  the above dynamical effects within the scale invariant vacuum (SIV)
theory and its  recent developments \cite{Maeder17a,Maeder23}.  This theory is resting on a most  simple and fundamental
question: is there still another fundamental symmetry in the sequence of Galilean invariance, Lorentz invariance  and Einstein general covariance ?
The case of scale invariance which is present in General Relativity and Maxwell equations for the empty space without charge and currents.
Cosmological models predicts a rapid decrease of these effects with increasing mean density up to the critical density where they vanish.
The  deep MOND limit,
where a gravity  $g= \sqrt{a_0 \, g_{\mathrm{N}}}$ (for Newtonian gravities $g_{\mathrm{N}} < a_0$),
appears as an  approximation of the SIV theory when the scale factor $\lambda$ can be considered as a constant. Such a condition 
is satisfied with deviations smaller than 1\%  for systems with dynamical timescales  $\leq$ 200 Myr. 
The scale invariant effects are generally  negligible in current life, they may however play a role in some 
astrophysical cases.
The SIV theory  already shows several positive results: on basic cosmological tests \citep{Maeder17a,MaedGueor20a,MaedGueor21a}
 with the occurrence of an accelerated expansion, 
on the growth of density fluctuations \cite{MaedGueor19}, 
as well  on  first dynamical tests \cite{Maeder17c}.

  Section 2 gives a presentation of the fundamental properties  and equations of the scale invariant theory.
  Section 3 is devoted to clusters of galaxies,  Section 4 to   rotation of galaxies  at the present time, as well as for redshifted galaxies.
  In Section 5,  the case of the velocity dispersion
  perpendicular to the galactic plane is examined.   Section 6 gives the conclusion.

  \section{Fundamental properties of the scale invariant theory}   \label{neuton}

Scale invariance is another symmetry in addition to the covariance of 
General Relativity. In this new symmetry, the equation of the gravitational
field allow possible changes of the length of the line element $ds$. 
  
 The scale invariant vacuum (SIV) theory tries to  follow the statement by Dirac
 \cite{Dirac73}:{ \emph{``It appears as one of the fundamental principles
in Nature that the equations expressing basic laws should be invariant
under the widest possible group of transformations''}}. Thus,  are  there further  invariances   to be considered in the theory of gravitation ?
 This simple question is very  meaningful since 
scale invariance is present in  General Relativity (GR)  in empty space without a cosmological constant
and is also present in Maxwell equations without charges and currents.  Thus, we may wonder whether  some   limited effects of scale invariance are 
still present in our very  low density Universe.
The point  is of interest regarding the long standing uncertainties  regarding the assumed  dark components.

The problem is not new. Weyl \cite{Weyl23} and Eddington \cite{Eddington23} considered scale invariance  in an attempt to account for
electromagnetism by a geometrical property of the space-time.  The proposition was then abandonned because Einstein \cite{Einstein18}
had shown that the properties of a particle would have depended on its past world line, so that 
 atoms  in an electromagnetic field would not show sharp lines.
Dirac \cite{Dirac73} and Canuto et al. \cite{Canuto77}
considered an interesting new possibility: the Weyl Integrable Geometry (WIG)  where Einstein's criticism does not apply as shown below.

\subsection{Basic assumptions}

  We refer to the above works by Dirac and Canuto et al.  for a more detailed study of the WIG space. In this Section 
 we just give a short summary of the main poperties of the SIV theory, see also \cite{Maeder17a,Maeder23,MaedGueor23}.
The WIG space is the appropriate framework  to study  scale invariance. Alike GR, it is endowed with a metrical connection
$ds^2 = g_{\mu\nu} dx^{\mu}dx^{\nu}$ and a line element $ds'$ of GR transforms into $ds$ in the WIG space,
\begin{equation}
 ds'= \lambda(x^{\mu}) \; ds, \quad \mathrm{with}  \; \lambda \mathrm{ \;the \; scale  \; factor.}
\label{ds}
\end{equation}
{\bf{At this stage, $\lambda$ remains unspecified, it will be defined below in Sect. \ref{ggg} by the choice of a physically  meaningful gauging condition.}}
Thus, there is a conformal relation $g'_{\mu \nu}= \lambda^2 g_{\mu \nu}$ between the two spaces.  Moreover, in the transport from a point 
$P_0(x^{\mu})$ to a nearby point  $P_1(x^{\mu}+dx^{\mu})$, the length
$\ell$ of a vector is assumed to change by
$ \;d\ell \, = \, \ell \, \kappa_{\nu} \, dx^{\nu}. $
There, $\kappa_{\nu}$ is called the coefficient of metrical connection, it  is a fundamental characteristics of the geometry
alike  $g_{\mu \nu}$. We note that in GR $\kappa'_{\nu}=0$.

The key difference between  Weyl's Geometry \citep{Weyl23} and the WIG is that in the latter $\kappa_{\nu}$ is the gradient of a scalar
field $\varphi$ \citep{Dirac73,Canuto77},
\begin{equation}
\kappa_{\nu}= -\varphi_{, \nu}=-\frac{\partial \ln \lambda}{dx^{\nu}}, \quad \quad  \mathrm{with}\;\; \varphi= \ln \lambda. 
\label{kak}
\end{equation}
This implies that $\kappa_{\nu}$ is a perfect differential with $\partial_{\nu} \kappa_{\mu}=\partial_{\mu} \kappa_{\nu}$ and thus 
 the parallel displacement of a vector along a closed loop in WIG does not change
its length.  Thus,   the length change of a vector does not depend on the path followed.
This is the  specificity of WIG \citep{Canuto77} with respect to the classical Weyl's Geometry \citep{Weyl23}
and  the criticism by Einstein does not apply in WIG. {\bfseries{This is equivalent to a scalar-tensor theory, however with a valuable property.
Scalar-tensor theories are characterized in general by the introduction of a scalar field $\varphi$, which is generally defined in an independent way, for some
specific purpose. A particularity of the scale invariant theory is that the scalar field $\varphi$ is determined by the scale factor $\lambda$, see Eq. (\ref{kak}).
Thus, there is no additional degree of freedom for chosing $\varphi$. The theory is more constraint without any adjustment parameter, a property 
that we consider as a quality of the theory in addition to the fact that an additional symmetry is considered.}}

 The inclusion of scale invariance demands further developments of the tensor calculus, with the introduction
of \emph{cotensors}. These mathematical  expressions enjoy  both the covariance of GR and the scale covariance. 
A quantity $Y,$ scalar, vector or tensor, which by a scale transformation
changes  like $Y\rightarrow\,Y' \,= \, \lambda^n(x) \, Y$, 
is said to be a {\emph{coscalar, covector, or cotensor }} of  power $\Pi(Y) =n$; one then speaks of {\emph{scale covariance}}.
For $n=0$ ({\emph{scale invariance}}), one has an {\emph{inscalar, invector, or intensor}}.
The ordinary derivatives  $Y_{,\mu}$  as well as  the covariant derivative $Y_{;\mu}$  are not necessarily
{\emph{co-covariant}}  (a definition by Dirac) or invariant,
but  derivatives  with such properties can be defined and are noted by $Y_{*\mu}$.

A brief summary
of co-tensor calculus has been given by Canuto et al. \cite{Canuto77}. Many useful
expressions can also be found in \cite{Dirac73}, as well as in \cite{BouvierM78}. 
 These developments are essential when studying  the scale invariant field equation in order to derive the 
cosmological equations  \cite{Maeder17a}. However at the level
of the geodesics we are considering here, we can manage without calling for cotensors.

\subsection{The  geodesics from an action principle}

The equation of geodesics is so  fundamental  that we provide herewith a detailed demonstration.  This equation  leads to: - the weak field or
Newtonian approximation, - the derivation of post-Newtonian effects, - and the lensing effect.
 In WIG, it can be obtained in different ways, as is the case in GR:

- A geodesic is a curve such that a co-vector tangent to the curve
is always transported by parallel displacement along the curve. In GR, one has 
$u'^{\mu}_{; \nu} u'^{\nu}=0$, and in WIG 
$u^{\mu}_{* \nu} u^{\nu}=0$ \cite{Canuto77}.

- It can be derived from the Equivalence Principle, which states \cite{Weinberg72} that  
at every point of the space-time there is a local inertial coordinate system $x'^{\alpha}$ such that 
$\frac{d^2 x'^{\alpha}}{ds'^2}=0$, a  condition applied by \cite{MBouvier79}.

- The geodesic can also be derived from an action principle \cite{Dirac73}.
Below, we give the  derivation  of this equation which is basic for the 
studies of motions in planetary systems and galaxies, as well as for the gravitational lensing \citep{BouvierM78}.

We could think to define the geodesics as the path along which the change of length of a vector
is minimum. This could apply to the classical Weyl's geometry, but not to WIG, since as said above, the
change of length during a displacement is independent of the path followed. Thus, we simply express
that the path $\mathcal{I}$ followed by a
particle   between two fixed points $P_0$ and $P_1$ should be an extremum 
and develop its implication   in the  WIG space  in terms of the two geometrical factors $g_{\mu \nu}$ and $\kappa_{\nu}$,  see Eq.~8.1 \citep{Dirac73}.
\begin{equation}
\mathcal{I} = \int^{P_1}_{P_0} ds' = \int^{P_1}_{P_0}   \lambda ds. 
\end{equation}
We express the following action principle,
\begin{equation}
 \delta \mathcal{I }= \delta  \int^{P_1}_{P_0}   \lambda ds=
 \int^{P_1}_{P_0}  \delta \left( \lambda {ds}\right) =\int^{P_1}_{P_0}  \delta \left( \lambda \frac{ds}{d\sigma}\right) \,d\sigma=  0\,,
\end{equation}
{\bfseries{where we have have introduced an affine parameter $\sigma$ in order to be able to take it outside the $\delta$ function.  
There $\sigma$ is a (co-)scalar like $s$; 
 the components of the tangent vector are $u^{\mu}=\frac{dx^{\mu}}{d\sigma}$, (equal to  $\frac{dx^{\mu}}{ds}$ if $\sigma=s$). We then have}}
\begin{equation}
\delta \left( \lambda \;\frac{ds}{d\sigma}\right)^2 = 2 \lambda \, \frac{ds}{d\sigma} \, \delta\left(\lambda \, \frac{ds}{d\sigma}\right).
\end{equation}
Thus, $\delta \mathcal{I}$ becomes,
\begin{equation}
\delta \mathcal{I}  = \int 
 \frac{1}{2 \lambda} \frac{d\sigma}{ds} \delta \left( \lambda^2 \; g_{\mu \nu} \frac{dx^{\mu}}{d\sigma} \frac{dx^{\nu}}{d\sigma}  \right) d\sigma=0.
\end{equation}
  Expressing the integrand, we get,
\begin{equation}
 \left[g_{\mu \nu} \, u^{\mu} u^{\nu} \delta \lambda+ 
\frac{\lambda}{2}\; \left(\frac{\partial}{\partial x^{\,\rho}}  g_{\mu \nu}\right)\; u^{\mu} u^{\nu} \,\delta x^{\,\rho}+
\lambda \, g_{\mu \nu} \, u^{\mu} \delta  u^{\nu}\right] \frac{d\sigma}{ds}.
\end{equation}
{\bfseries{In the last term, we can write
$\delta \, u^{\nu}= \delta\left( \frac{dx^{\nu}}{d\sigma} \right)= \frac{d}{d\sigma} (\delta x^{\nu})$ and the last term can be integrated by parts.
One also has $\delta x^{\nu}(P_0)=0$ and the same for $P_1$,  and  $\delta I$ becomes,}}
\begin{equation}
\delta \mathcal{I} = \int \left[g_{\mu \nu}  \frac{\partial \lambda}{\partial x^{\, \rho}} \,\ u^{\mu} u^{\nu}\frac{d\sigma}{ds}+
\frac{1}{2}  \lambda \left(\frac{\partial g_{\mu \nu}  }{\partial x^{\, \rho}}\right) \, u^{\mu} u^{\nu}\frac{d\sigma}{ds}-
\frac{d}{d\sigma } \left(g_{\, \rho \mu} \, u^{\mu} \lambda  \frac{d\sigma}{ds}\right) \right] \delta x^{\rho} d\sigma=0 \, .
\end{equation}
Now we can develop the last term in the integrand,
\begin{equation}
- \frac{d}{d\sigma } \left(g_{\, \rho \mu} \, u^{\mu} \lambda  \frac{d\sigma}{ds}\right)= -\left(\frac{\partial}{\partial x^{\nu}}g_{\rho \mu}\right)  
  \lambda u^{\nu} u^{\mu}\frac{d\sigma}{ds}  -\lambda g_{\rho \mu} \frac{d u^{\mu}}{d\sigma} \frac{d \sigma}{ds}
-\left(\frac{\partial \lambda}{\partial x^{\nu}}\right) g_{\rho \mu} \, u^{\nu} u^{\mu} \frac{d\sigma}{ds}\,.
\end{equation}
{\bfseries{There is a last term $-g_{\rho \mu}u^{\mu} \lambda \frac{d}{d \sigma}(\frac{d\sigma}{ds})$, which  disappears
since $\sigma$ is proportional to $s$, and in particular if $\sigma=s$.  
Thus, the constant ratio  $\frac{d \sigma}{ds}$, present in each term of the 
integral  can be removed and the  action principle now becomes,}}
\begin{eqnarray}
\quad \quad \quad \delta \mathcal{I} &=& \int \left[ g_{\mu \nu}  \frac{\partial \lambda}{\partial x^{\, \rho}} \,\ u^{\mu} u^{\nu}+
\frac{1}{2}  \lambda \left(\frac{\partial g_{\mu \nu}  }{\partial x^{\, \rho}}\right) \, u^{\mu} u^{\nu} \right]  \delta x^{\,\rho} d\sigma \nonumber    \\
 &-& \int \left[\lambda \frac{\partial g_{\,\rho \mu}}{\partial x^{\nu}}u^{\mu} u^{\nu}+\lambda g_{\,\rho \mu} \frac{du^{\mu}}{d\sigma}  
+ g_{\,\rho \mu} \frac{\partial \lambda}{\partial x^{\nu}}u^{\mu} u^{\nu} \right]  \delta x^{\,\rho} d\sigma =0 \,.
\end{eqnarray}
This applies   for any displacements $\delta x^{\,\rho}$   along the curve,
 thus the integrand vanishes at all points. Expressing the metrical connection  {\bfseries{according to Eq.(\ref{kak})}}
	$\kappa_{\nu} =- \frac{1}{\lambda}\frac{ \partial \lambda}{\partial x^{\nu}}$, we get
\begin{equation}
\quad g_{\,\rho \mu} \frac{du^{\mu}}{d\sigma}+\left(-\frac{1}{2} \frac{\partial g_{\mu \nu}  }{\partial x^{\, \rho}}+
  \frac{\partial g_{\,\rho \mu}}{\partial x^{\nu}}+g_{\mu \nu} \kappa_{\,\rho}- g_{\,\rho \mu} \kappa_{\nu}\right) u^{\mu} u^{\nu}  =0.
\end{equation}
We can write $\frac{\partial g_{\,\rho \mu}}{\partial x^{\nu}}u^{\mu} u^{\nu}= 
\frac{1}{2} \left(\frac{\partial g_{\,\rho \mu}}{\partial x^{\nu}}+\frac{\partial g_{\,\nu \rho}}{\partial x^{\mu}}\right) u^{\mu} u^{\nu}$
and obtain,
\begin{equation}
\quad g_{\,\rho \mu} \frac{du^{\mu}}{d\sigma}+ \Gamma_{\rho,\mu \nu}\, u^{\mu} u^{\nu}+
 \kappa_{\,\rho} \,g_{\mu \nu}u^{\mu} u^{\nu}-\kappa_{\nu} g_{\,\rho \mu} u^{\mu} u^{\nu}=0,
\end{equation}
Multiplying this equation by $g^{\,\rho \alpha}$, one obtains
\begin{equation}
\quad  \frac{du^{\alpha}}{d\sigma}+ \Gamma^{\alpha}_{\mu \nu}\, u^{\mu} u^{\nu}+
 \kappa^{\alpha} \,g_{\mu \nu}u^{\mu} u^{\nu}-\kappa_{\nu}  u^{\nu} u^{\alpha}=0.
\end{equation}
Using the above expressions of  $u^{\mu}$ and $u^{\nu}$  when taking $\sigma=s$, we get simply
\begin{equation}
\frac{du^{\alpha}}{ds}+ \Gamma^{\alpha}_{\mu \nu}\, u^{\mu} u^{\nu}
 -\kappa_{\nu}  u^{\nu} u^{\alpha}+\kappa^{\alpha} =0,
\label{GEOD}
\end{equation}
\begin{equation}
\mathrm{with} \quad \Gamma^{\alpha}_{\mu \nu}= g^{\rho \alpha} \Gamma_{\rho,\mu \nu}  \quad \mathrm{with} \; \;
\Gamma_{\rho,\mu \nu} = \frac{1}{2} \left(\frac{\partial g _{\mu \,\rho}}{\partial x^{\nu}  }+
\frac{\partial g_{\nu \, \rho }}{\partial x^{\mu} }-\frac{\partial g_{\mu \nu}  }{\partial x^{\,\rho  }}\right),  
\end{equation}
and
\begin{equation}
\Gamma^{\alpha}_{\mu \nu} = \frac{1}{2} g^{\sigma \alpha} \left(\frac{\partial g _{\nu \,\sigma}}{\partial x^{\mu}  }+
\frac{\partial g_{\mu \, \sigma }}{\partial x^{\nu} }-\frac{\partial g_{\nu \mu}  }{\partial x^{\,\sigma  }}\right)\, .
\end{equation}
Eq.~(\ref{GEOD}) only depends on the WIG geometrical factors $g_{\nu \mu}$ and $\kappa_{\nu}$.
It is the same expression as that obtained by the other  methods. 
If $\lambda$ is a constant, one has $\kappa_{\nu}=0$ and  the geodetic equation is the same as the classical one, cf. \cite{Weinberg72}.

\subsection{The Newtonian approximation in weak fields} \label{Isaac}

The Newtonian approximation was obtained {\bfseries{from the above geodesic equation}} by \cite{MBouvier79} for a weak stationary potential $\Phi$, with 
a line element only slightly differing from Minkowski's, with
\begin{equation}
g_{ii} \cong -1, \quad  i=1,2,3 \quad \mathrm{and} \; \; g_{00}= 1+ 2 \, { \Phi}\,,
\label{Phi}
\end{equation}
 setting  $c=1$ and the same below.
 With the above $g_{\mu\nu}$, the $\Gamma^{\alpha}_{\mu \nu}$ vanish,
except $\Gamma^{i}_{00}= g^{i \alpha} \frac{\partial \Phi}{\partial x^{\alpha}}$ and for
a potential of the classical form $\Phi= GM/r$, this just gives $\Gamma^{i}_{00}=-  \frac{GM}{r^2}$
For slow motions vs. $c$, $u^i \rightarrow \upsilon^i= \frac{dx^i}{dt}$ and $u^0 \rightarrow 1$, with $ds \rightarrow dt$.
  All these are just the approximations currently made in the weak field low velocity approximation \citep{Weinberg72}.
At this stage, the expressions  of $\lambda$ and thus of $\kappa_{\nu}$ are not specified, this will be done in Sect. \ref{ggg}.
For now, let us consider the case of the empty space  with $\kappa_{\nu} =(0, 0, 0, \kappa_0)$.  
Eq.~(\ref{GEOD}) is leading to,
\begin{equation}
 \frac{d\upsilon^i}{dt}+  g^{i \alpha} \frac{\partial \Phi}{\partial x^{\alpha}}  - \kappa_0 \upsilon^i =0,
\end{equation}
and in vectorial form,
\begin{equation}
 \frac{d^2 \vec{r}}{ dt^2}= - \frac{GM}{r^2} \frac{\vec{r}}{r} +\kappa_0 \frac{d \vec{r}}{dt}\,.
\label{Nvec}
\end{equation} 
An acceleration term proportional to  the velocity and in the same direction appears in the 2$^{nd}$ member.
It is such as to accelerate an expansion or a contraction depending on circumstances. Such  effects  appear {\emph{e.g.}} in the cosmological
models \citep{Maeder17a} and in   the rapid growth of density fluctuations in the early Universe \citep{MaedGueor19}.

Some properties of the theory, such as $\lambda$, $\kappa_{\nu}$ and  the way the differents physical quantities depend on the 
scale $\lambda$, are determined by the conservation laws and the considered cosmological models. Therefore, it is now necessary to specify a few
physical properties  in presence of a varying gauge.

\subsection{The scale invariant field equation and the gauge}  \label{ggg}

For detailed developments of the scale invariant theory and cosmology, the reader may refer to \cite{Maeder17a,MaedGueor20a,Maeder23}.
Here we limit the presentation to a few relevant points.	A general scale invariant field equation was obtained
by \cite{Canuto77},
\begin{eqnarray}
R'_{\mu \nu} - \frac{1}{2} \ g_{\mu \nu} R'-\kappa_{\mu ,\nu}-\kappa_{ \nu ,\mu}
-2 \kappa_{\mu} \kappa_ {\nu} 
+ 2 g_{\mu \nu} \kappa^{ \alpha}_{,\alpha}
- g_{\mu \nu}\kappa^{ \alpha} \kappa_{ \alpha} = 
-8 \pi G T_{\mu \nu} - \lambda^2 \Lambda_{\mathrm{E}} \, g_{\mu \nu}, \, 
\label{field}
\end{eqnarray}
\noindent
where $R'_{\mu \nu}$ is the Ricci tensor in GR and $R'$ its contracted form, $ \Lambda_{\mathrm{E}}$ the cosmological constant 
in GR, $G$ is  a true constant.  A fully  detailed demonstration of this field equation by an action principle has also been recently established
\cite{MaedGueor23} in  the line of earlier works by \cite{Dirac73},  who established the expression of the first member 
which is  scale invariant, and by  \cite{Canuto77} who also further developed the WIG concepts.

 The second member of the field equation must also be scale invariant, and
 the  energy-momentum tensor $ G   \, T_{\mu \nu}$ 
 has the same property.  Thus,  $T_{\mu \nu} \,= \,T '_{\mu \nu}$,
which has  implications on  the relevant densities,  \emph{e.g.} for a perfect fluid,
\begin{equation}
( p+\varrho) u_{\mu} u_{\nu} -g_{\mu \nu } p =
( p'+\varrho') u'_{\mu} u'_{\nu} -g'_{\mu \nu } p' \, .
\end{equation}
The  velocities $u'^{\mu}$ and $u'_{\mu}$ transform like,
\begin{eqnarray}
u'^{\mu}&=&\frac{dx^{\mu}}{ds'}=\lambda^{-1} \frac{dx^{\mu}}{ds}= \lambda^{-1} u^{\mu} \, , \nonumber \\
\mathrm{and} \; \;
u'_{\mu}&=&g'_{\mu \nu} u'^{\nu}=\lambda^2 g_{\mu \nu} \lambda^{-1} u^{\nu} = \lambda \, u_{\mu} \, .
\label{pl1}
\end{eqnarray}
The contravariant and covariant components of a vector have different powers and
 $T_{\mu \nu}$ is scaling like, 
\begin{equation}
( p+\varrho) u_{\mu} u_{\nu} -g_{\mu \nu } p =
( p'+\varrho') \lambda^2 u_{\mu} u_{\nu} - \lambda^2 g_{\mu \nu } p' \, ,
\end{equation}
 implying   for $p$ and $\varrho$ \citep{Canuto77},
\begin{equation}
p = p' \, \lambda^2 \, \quad \mathrm{and} \quad \varrho = \varrho' \, \lambda^2 \, .
\label{ro2}
\end{equation}
The consistency of the field equation implies that $p$ and $\varrho$ are not scale invariant, but coscalars of power $\Pi(\rho)=-2$.
These properties  are consistent with  the equation of conservation  (\ref{conserv}) discussed below.

Chosing a metric is necessary to fix the $g_{\mu \nu}$, similarly a gauging condition is necessary to fix the scale
or gauge $\lambda$. Dirac  \cite{Dirac73} and Canuto et al. \cite{Canuto77}  had chosen the  so-called {\it{``Large
Number Hypothesis``}}, see also \cite{Dirac74}. 
The author's choice is to adopt   the following statement \cite{Maeder17a}:
  {\emph{The macroscopic empty space is scale invariant, homogeneous and isotropic}}. 
  The  equation of state of the vacuum $p_{\mathrm{vac}}= -\varrho_{\mathrm{vac}}  c^2$  is precisely 
the relationship permitting  the vacuum density   to remain constant for an adiabatic
expansion or contraction  \cite{Carroll92}. The above gauging condition also implies $\lambda=\lambda(t)$, 
 since the empty space is homogeneous and isotropic.  Applied to the field 
equation in empty space, the gauging condition is leading to two differential equation for $\lambda$ which are then
 only a function of time. {\bfseries{The complete details of this derivation are given by Maeder \cite{Maeder17a}}},
\begin{eqnarray}
\  3 \, \frac{ \dot{\lambda}^2}{\lambda^2} \, =\, \lambda^2 \,\Lambda_{\mathrm{E}}  \,  
\quad \mathrm{and} \quad  \,  2\frac{\ddot{\lambda}}{\lambda} - \frac{ \dot{\lambda}^2}{\lambda^2} \, =
\, \lambda^2 \,\Lambda_{\mathrm{E}}\,,
\label{diff1}
\end{eqnarray}
Impressively enough, the trace of Eq.~(\ref{field}) in empty space is also leading to these two equations \cite{MaedGueor23}.
Their solution is, 
\begin{equation}
\lambda(t) =\sqrt{\frac{3}{\Lambda_{\mathrm{E}}}}  \frac{1}{ct} \,    \quad \mathrm{or} \; \; \lambda(t)= \frac{t}{t_0},
\label{lambda}
\end{equation}
with $\lambda=1$ for $t_0=1$. This implies that  the only component of $\kappa_{\nu}$ is $ \kappa_0= - \dot{\lambda}/\lambda =1/t.$
 Thus, we see that the above Eq.~(\ref{Nvec}) has consistent unities.
The form of Eqs.~(\ref{diff1}) and (\ref{lambda}) is independent of the matter content according to the gauging condition, but the range
of variations of $t$, and thus of $\lambda$ and $\kappa_{\nu}$,  are set by the considered comological models \cite{Maeder17a}, {\bfseries{see also
remarks following Eq. (\ref{Jesus})}}.
	
\subsection{Cosmological constraints}	

The application of the  FLWR  metric to Eq.~(\ref{field}) led  to rather "heavy"
 cosmological  equations  \cite{Canuto77},  however the  equations~(\ref{diff1})
bring  interesting simplifications, letting only one term with a $\lambda$-function
\cite{Maeder17a},
\begin{eqnarray}
\frac{8 \,\pi G \varrho }{3}\, &=&\, \frac{k}{a^2}+\frac{\dot{a}^2}{a^2}+ 2 \,\frac{\dot{a} \dot{\lambda}}{a \lambda}, \label{E1} \\
-8 \, \pi G p \, &=& \,\frac{k}{a^2}+ 2 \frac{\ddot{a}}{a}+\frac{\dot{a^2}}{a^2}
+ 4 \frac{\dot{a} \dot{\lambda}}{a \lambda} ,\label{E2} \\
- \frac{4\pi G}{3} \left(3p+\varrho \right) \,&=& \,\frac{\ddot{a}}{a} + \frac{\dot{a} \dot{\lambda}}{a \lambda}. \label{E3}
\end{eqnarray}
\noindent
These  equations  differ from Friedmann's by only one term. For a constant $\lambda$, 
 Friedmann's equations are recovered.
The 3$^{rd}$ equation is  derived from the first two.  
Since $\dot{\lambda}/{ \lambda}$ is negative,  the extra term represents  an additional acceleration {\emph{ in the direction of  
motion}} and  is thus fundamentally different from that of the cosmological constant.

Analytical solutions  for the flat SIV models with $k=0$   were found
for the  matter  era by \cite{Jesus18},
\begin{equation}
a(t) =  \left[\frac{t^3 -\Omega_{\mathrm{m}}}{1 - \Omega_{\mathrm{m}}} \right]^{2/3}, \quad \quad \mathrm{the\; initial\; time\;is} \; \;
t_{\mathrm{in}}  = \Omega^{1/3}_{\mathrm{m}},
\label{Jesus}
\end{equation}
\noindent
and for radiation  dominated  models  by \cite{Maeder19}. These equations  allow flatness
for different values of  $\Omega_{\mathrm{m}}$, unlike Friedmann's models.
At  present  $t_0=1$ and  $a(t_0)=1$.
The graphical solutions are relatively  close to the corresponding $\Lambda$CDM models \citep{Maeder17a},
with larger differences for very low $\Omega_{\mathrm{m}}$.
The initial time  $ t_{\mathrm{in}}$ of the origin is defined by $a( t_{\mathrm{in}})=0$. 
For $\Omega_{\mathrm{m}}=0, 0.01, 0.1, 0.3, 0.5$, the values of  $ t_{\mathrm{in}}$ are 0, 0.215, 0.464, 0.669, 0.794 respectively.
This strongly reduces the possible range of $\lambda(t)= (t_0/t)$,
between $1/t_{\mathrm{in}}$ and $1/t_0=1$.  {\it{Scale invariant effects are drastically reduced in Universe models 
with increasing $\Omega_{\mathrm{m}}$}}. They are killed
for  $\Omega_{\mathrm{m}} \geq 1$.

While   $t$  varies  between $t_{\mathrm{in}}$  and
$t_0=1$ at  present;
the usual timescale $\tau$  in years or seconds   varies  from $\tau_{\mathrm{in }}=0$   and
$\tau_0= 13.8$ Gyr at  present  \citep{Frieman08}.
The  relation between these  age units    is
$\frac{\tau - \tau_{\mathrm{in}}}{\tau_0 - \tau_{\mathrm{in}}} = \frac{t - t_{\mathrm{in}}}{t_0 - t_{\mathrm{in}}}$,
expressing that the age fractions of an event  are the same.
This gives
\begin{eqnarray}
\tau \,= \, \tau_0 \, \frac{t- t_{\mathrm{in}}}{1- t_{\mathrm{in}}} \,  \quad \mathrm{and} \; \;
t \,= \, t_{\mathrm{in}} + \frac{\tau}{\tau_0} (1- t_{\mathrm{in}}) \,,\label{T2} \\
\mathrm{with} \quad \frac{d\tau}{dt} \, = \, \frac{\tau_0}{1-t_{\mathrm{in}}}\,, \quad \mathrm{and}\; \;
\frac{dt}{d\tau} \, = \, \frac{1-t_{\mathrm{in}}}{\tau_0}\,.   \label{dT1}
\end{eqnarray}
Both timescales are evidently linearly related
  $\frac{dX}{d\tau}= \frac{dX}{dt} \frac{dt}{d\tau}$, with a constant $\frac{dt}{d\tau}$ for a given $\Omega_{\mathrm{m}}$.

Eqs.~(\ref{E1}) and (\ref{E2})  lead to the scale invariant mass-energy conservation law  \citep{Canuto77,Maeder17a},
\begin{eqnarray}
\frac{d(\varrho a^3)}{da}+ 3 \, p a^2 +(\varrho+3p) \frac{a^3}{\lambda} \frac{d\lambda}{da}=0, \\
\mathrm{and} \quad \varrho \, a^{3(1+w)} \lambda^{1+3w} = \mathrm{const.}  \quad \mathrm{with\; p=w \varrho}, \label{conserv}
\end{eqnarray}
(there $c=1$). For ordinary matter $w=0$ and $\varrho_{\mathrm{m}} a^3 \lambda= const.$ A length element  
scales like $\lambda^{-1}$, thus from (\ref{conserv}) the  mass also varies like $\lambda^{-1} \sim t$. 
Such a mass variation might at first be surprising, however
let us recall that the invariance to Lorentz transformations, assumed in Special Relativity, is also leading to the concept of non-constant 
masses, which may vary up to infinity as a function of the velocities. Here, as in Special Relativity
 the number of particles remains the same, but their inertial and gravitational properties may  vary.
This property  is quite consistent with the above mentioned scale invariance of the
energy momentum tensor $T_{\mu \nu}$  in the second member of the general field equation (\ref{field}).
It implies that $\varrho = \varrho'  \lambda^{2}$ (cf. Eq.~\ref{ro2})  or  $\frac{M}{R^3} = \lambda^2 \frac{M'}{R'^3}$,
since $R'= \lambda R$ one has $M = M'/\lambda$, meaning  that $M \sim t$.
 A noticeable consequence of the mass variation is that the gravitational potential $\Phi = - GM/R$ 
of a given mass appears as a more fundamental quantity than the mass, being scale invariant and thus independent of time through the
evolution of the Universe. 
These results about an additional physical invariance
 are consistently derived from first principles, however this does not 
prove that Nature is proceeding this way. The purpose of the present work is precisely to proceed to some
observational tests.

The mass changes are  very limited,  from (\ref{T2}) they behave like
$M(\tau) = M(\tau_0) [t_{\mathrm{in}} + \frac{\tau}{\tau_0} (1- t_{\mathrm{in}})]$. 
As examples for $\Omega_{\mathrm{m}}=0.30$, the initial mass of the Sun was 0.891 of the present one and
 400 Myr ago, masses were differing by less than 1\% from today's. Thus, galactic masses are about constant 
over such a time interval and in this case (only) the deep MOND limit appears  as an approximation of the scale invariant theory \citep{Maeder23}.


Equation~(\ref{Nvec}) was expressed in variable $t$, where the age of the Universe is  $(1- t_{\mathrm{in}})$.
One has $\frac{d^2 r}{dt^2} =\frac{d^2 r}{d \tau^2} (\frac{d\tau}{dt})^2$, and
the constant $G$   in $t$-units becomes
$G(\frac{dt}{d\tau})^2 \; (\equiv G$  in current units). 
Thus at  present  $\tau_0$, one has
\begin{equation}
\frac {d^2 \bf{r}}{d \tau^2}  = - \frac{G  M }{r^2}  \frac{\bf{r}}{r}   + \frac{\psi_0}{\tau_0}   \frac{d\bf{r}}{d\tau} ,
\quad \mathrm{with} \; \; \psi_0 =1-t_{\mathrm{in}} \,.
\label{Nvec4}
\end{equation}
\noindent
For $\Omega_{\mathrm{m}} =0,  0.05, 0.10, 0.20, 0.30, 1$,  one has $\psi_0=1, 0.632, 0.536, 0.415, 0.331, 0.$
At other epochs $\tau$,  instead of $\psi_0$ one has the numerical factor 
\begin{equation}
\psi(\tau) = \frac{t_0-t_\mathrm{in}}{\left( t_\mathrm{in} +   \frac{\tau}{\tau_0} (t_0-t_\mathrm{in})\right) }, \quad \quad
 \left(  = \frac{t_0-t_{\mathrm{in}}}{t}  \; \right).  \label{pss}
\end{equation}
The  additional acceleration, {\it{the dynamical gravity}},
 is   generally  small since $\tau_0$ is  large. In an  empty universe, $\psi_0= 1$ and the dynamical acceleration is the largest,  
while  for  $\Omega_{\mathrm{m}} \geq 1$, $\psi_0=0$ and the additional term consistently disappears.

\section{Dynamics of  the clusters of galaxies}  \label{clust}

The above prescriptions are now submitted to major observational tests, starting with the velocities in clusters of galaxies.

\subsection{Observational constraints}

Since the historical discovery by \cite{Zwicky33},  later followed by \cite{Karachantsev66}, 
of large mass excesses in clusters of galaxies when comparing galaxy velocities 
and luminosities, evidences have further accumulated  (e.g. \cite{Blindert04}, \cite{Proctor15}, \cite{Sohn17}). 
Virial masses in excess with respect to stellar luminosities
 have been confirmed, reaching a factor of about 30-50   \cite{Proctor15} for large clusters 
with masses between $10^{14}$ and $10^{15}$ M$_{\odot}$. 
This implies that the stellar mass fraction is only of the order of 0.02-0.03.
The mass estimates can be affected by uncertainties in the mass-luminosity ratio $(M/L)$
 and also by different dynamical substructures within clusters
 \cite{Old18}. However, the above quantitative estimates  are  unchanged. 
As mentioned above, there is a relation between the total dynamical
 mass of clusters and  the baryonic masses of central regions   \cite{Chan22}. 
Such an unexpected relation, also found by \cite{Lelli17} in galaxies,  is challenging for the current $\Lambda$CDM models.

  X-ray observations of the emitting gas  allow estimates of  the gas mass  fraction (0.10-0.15) in clusters
   \cite{Lin12,Leauthaud12,Gonzalez13,Shan15,Ge16,Chiu16}. This largely dominates
  over the stellar mass fraction mentioned above.  These two fractions contribute to the baryon mass fraction,
  $f_{\mathrm{bar}} = \frac{M_{\mathrm{star}}+M_{\mathrm{gas}}}{M_{\mathrm{tot}}}$. 
   The total baryon fraction  appeared nearly constant with cluster mass  between 0.12 and 0.15 \cite{Gonzalez13}. 
   From a sample of 120 clusters, a gas mass fraction of 
  0.156($\pm0.003$) was estimated  \cite{Planck20}. 
   From a sample of 40 clusters with deep Chandra, 
  X-ray observations determined \cite{Mantz22} a gas mass fraction on the average similar, but with a  trend to increase with
  the overdensity $\Delta$ (the ratio to the critical density at the considered redshift). 
  In the Perseus cluster, the gas mass fractions are in the range of 0.09 to 0.20 for $\Delta$ going from $10^4$
  to 500. Gas mass fractions between 0.12 and 0.16
  were  found  \cite{Wicker23}, with
  increasing  values with redshift $z$, also increasing
  as a function of their masses, an   effect possibly due to a bias due
   the hypothesis of hydrostatic equilibrium in  cluster models.
 
  Summing the stellar and gas mass fractions one gets the baryon mass fraction which turns around 0.15 - 0.18, a factor
  of about  6 smaller than unity, a result in agreement with \cite{Planck20} 
  for matter, baryon and dark   matter density parameters: 
  $\Omega_{\mathrm{m}}=0.315 \pm 0.007$, 
 $\Omega_{\mathrm{b}}=0.050 \pm 0.0002$ and  
 $\Omega_{\mathrm{DM}}
 =0.265 \pm 0.002$.

\subsection{Cluster dynamics in the scale invariant context}

  This matter  was already considered \cite{Maeder17c}, however an update to include  
the limited  $\lambda$-variations and new observations  is necessary. This reduction of the scale invariant effects is expressed by the factor
  $\psi_0$ (or $\psi(\tau)$) in the equation of motion (\ref{Nvec4}),  as well as in its various applications.
  The  question is now whether in the scale invariant context the baryonic 
  matter alone is sufficient or not to account for the 
  high observed velocities in galaxy clusters.

  As the current
Virial theorem does not apply in an expanding system,  we start from the equation of motion (\ref{Nvec4}) and consider 
 a galaxy $i$  with velocity $\upsilon_i$ attracted  by 
  another $j$-one at time $\tau$ and distance $r_{ij}$. Its acceleration is
 \begin{eqnarray}
 \frac{d\upsilon_i}{d \tau} = \frac{G  m_j(\tau)}{r^2_{ij}}    + \frac{\psi(\tau)}{\tau_0}   \upsilon_i ,  \quad \mathrm{with} \; \;
\psi(\tau)  = \frac{1}{\frac{\Omega^{1/3}_{\mathrm{m}}}{1-\Omega^{1/3}_{\mathrm{m}}}+ \frac{\tau}{\tau_0} } .  \label{Nv1}\\
\mathrm{thus} \quad \frac{\psi(\tau)}{\tau_0} = \frac{1}{\tau_{\Omega}+\tau}\, , \quad \mathrm{with} \; \; 
\tau_{\Omega}=\tau _0 \,\frac{\Omega^{1/3}_{\mathrm{m}}}{1-\Omega^{1/3}_{\mathrm{m}}},
 \label{Nvecv}
\end{eqnarray}
\noindent
where  $ \Omega^{1/3}_{\mathrm{m}}= t_{\mathrm{in}}$, the initial time (Big Bang).   
 We  multiply the first equation by $\upsilon_i = dr_{ij}/d\tau$, integrate and sum  on all the other $(N-1)$  $m_j$-masses.
The variations of the mass $m_j(\tau)$ are  very limited, thus we consider a mean value  $ p \, m_j(\tau_0)$ 
of  masses $m_j$ at the middle age of the Universe,  thus $p$ is given by
 $p \approx (1+\Omega^{1/3}_{\mathrm{m}})/2$. Values of $p$ for different $ \Omega_{\mathrm{m}}$
 are given in Table 1 (the range of the $p$-values is small). One is getting,
 \begin{equation}
\frac{1}{2} \upsilon^2_i \, = \,  p  \sum_{j\neq i} \frac{G\,  m_j (\tau_0)} {r_{ij}} + 
\int  \frac{1}{\left[\tau_{\Omega}+\tau \right]} \, \upsilon^2_{i} d\tau \, .
\label{en1}
\end{equation} 
 \noindent
Then, we are also summing on all the N  $i$-masses and get,
 \begin{equation}
\frac{1}{2}\sum_i \upsilon^2_i  = \frac{1}{2}  p \sum_i \sum_{j\neq i} \frac{G\,  m_j (\tau_0)} {r_{ij}} + 
\sum_i\int  \frac{1}{\left[\tau_{\Omega}+
\tau \right]} \, \upsilon^2_{i} d\tau \, .
\label{en1}
\end{equation} 
 \noindent
  The factor $1/2$ in front of the double summation is necessary 
 in order not to   account twice the same masses.

\begin{table*}
\vspace*{0mm} 
 \caption{ Parameters for cluster dynamics and  ratios of the mass estimates 
in the standard case and in the scale invariant theory, calculated with $\tau_0= 13.8$ Gyr  and $\tau_1 = 400 $ Myr.  }
\begin{center} 
\scriptsize
\begin{tabular}{ccccccccccc}
$\Omega_{\mathrm{m}}$  &  $p$  & $\tau_{\Omega}$ & $x(\Omega _{\mathrm{m}})$  &  
$M_{\mathrm{Std}}/M_{\mathrm{SIV}}$   & $M_{\mathrm{Std}}/M_{\mathrm{SIV}}$   &$M_{\mathrm{Std}}/M_{\mathrm{SIV}}$  & 
 $M_{\mathrm{Std}}/M_{\mathrm{SIV}}$  &  $M_{\mathrm{Std}}/M_{\mathrm{SIV}}$ & $M_{\mathrm{Std}}/M_{\mathrm{SIV}}$&  \\
  &   &[Gyr]  &   &  $f= 0.5$  &  $f=0.6$   & $  f=0.7$&$  f=0.8$&$  f=0.9$&$  f=1.0$&\\
\hline
 &   &   &   \\
0.05  & 0.684 &  8.05 & 0.950 & 13.68     & $\infty$   &$\infty$ & $\infty$ &$\infty$ &$\infty$ &\\
0.10  & 0.732 & 11.95 & 0.735 & 2.76      &      6.20      &$\infty$ &$\infty$ &$\infty$ &$\infty$ & \\
0.20  & 0.792 & 19.44 & 0.516 & 1.64      &      2.07       &   2.85  &$4.54 $ &$11.12$ &$\infty$ &\\
0.30  & 0.835 & 27.95 & 0.387 & 1.36      &      1.56       &   1.82  & $2.19$ &$2.75$ &$3.69$ & \\
\hline
\end{tabular}
\end{center}
\footnotesize{Note: the signs $\infty$ indicate a divergence of the ratio   $M_{\mathrm{Std}}/M_{\mathrm{SIV}}$ in Eq.~(\ref{mratio})   }.
\normalsize
\end{table*}

 Now, we divide the above equation by $N$.
  The   term on the left becomes
  $\frac{1}{2} \overline{\upsilon^2}$.  The first term on the right gives the potential energy
 $ p q G M/R$ of the cluster with a total present mass   $M$ and a radius $R$, with $q$
  a structural factor depending on the density profile ($q=3/5$ for an homogeneous distribution).
   The last
 term is an adiabatic invariant, for which we take an average over the lifetime of the cluster
  between $\tau_1$ and $\tau_0$, where $\tau_1$ is the formation epoch.  It is then  easy to integrate it,
 \begin{eqnarray}
  f \overline{\upsilon^2}\, \int ^{\tau_0}_{\tau_{1}}  \frac{ d\tau}{\left[\tau_{\Omega}+\tau\right]}\,= 
  f\overline{ \upsilon^2} \,\ln  \frac{\tau_{\Omega}+\tau_0}{\tau_{\Omega}+\tau_1} = 
  f \overline{\upsilon^2}  \times  x(\Omega_{\mathrm{m}}), \\
   \mathrm{with} \; \; x(\Omega_{\mathrm{m}})= 
  \ln  \frac{\tau_{\Omega}+\tau_0}{\tau_{\Omega}+\tau_1}. \quad  \quad
 \label{t3}
  \end {eqnarray}
\noindent
The   factor $f$ accounts  for the integration of $\upsilon^2$ over time $\tau$; different values of $f$
between 0.5 and 1.0 are considered  in Table 1. The values of $ x(\Omega_{\mathrm{m}})$ as a 
function of different $\Omega_{\mathrm{m}}$ are also given in the table.
The integration constant on the first term in (\ref{en1})
may be considered as zero, since a zero velocity  is an appropriate value for the end of the collapse; 
the same for the potential which is vanishing at large distances.
Equation~(\ref{en1})  can be written as,
\begin{eqnarray}
\overline{\upsilon^2} \, = \,  2 \,\frac{ pq \,G M_{\mathrm{SIV}}}{R}+
 2  \, x(\Omega_{\mathrm{m}}) \times f \overline{\upsilon^2} \,,
\label{en22}
\end{eqnarray}
\noindent
where $M_{\mathrm{SIV}}$ is the present mass in the scale invariant theory.
In  the standard case,  a value $ M_{\mathrm{std}}$ 
would be derived from the same value of the velocity dispersion  $\overline{\upsilon^2} $
 by  the relation,
 \begin{equation}
   \overline{\upsilon^2} \, = \,  2 \, q \,G M_{\mathrm{std}}/{R}.
   \label{stdratio}
   \end{equation} 
 The ratio of the two  mass estimates is given by,
 \begin{equation}
\frac{M_{\mathrm{std}}}{M_{SIV}} \,  \approx \, \frac{p}{ \left[1 - 2 \, f \,x(\Omega_{\mathrm{m}}) \right]}.
\label{mratio}
\end{equation}
The same structural factor $q$ is taken in the standard and scale invariant cases.
 Also, we may consider that the statistical factors,  which account for the fact that
the observed velocities are the radial ones, are the same in the two theories.
 The factors $x(\Omega_{\mathrm{m}})$ are estimated by Eq.~(\ref{t3}),
  taking a ratio $\tau_1/\tau_0= 3\%$, 
 which corresponds to  galaxy formation at a cosmic time of 400 Myr,
 the choice of the initial time being not critical. We note that the factor $p$ has no effect in the term responsible of the divergence.

  Table 1 gives the ratios of the mass estimates $\frac{M_{\mathrm{std}}}{M_{SIV}}$.
  We see that for the whole range of $\Omega_{\mathrm{m}}$ values, the standard
  mass estimates are much larger than the SIV ones. The infinite values in the table result from the vanishing
  denominators of Eq.~(\ref{mratio}) for the mass ratios $\frac{M_{\mathrm{std}}}{M_{SIV}}$.
 Particularly, for low values of $\Omega_{\mathrm{m}}$
  such  as $\Omega_{\mathrm{m}}=0.05$, the  overestimate of 
  the masses becomes very important. Thus,  if scale invariant effects are present in Nature, we may expect,  by applying  standard estimates,
 large overestimates of the real masses of the galaxy clusters, with then the  necessity of advocating 
large amounts of dark matter.
  Of course, in the process of standard determinations from an expression like Eq.~(\ref{stdratio})
   no mathematical divergence is occurring.  Nevertheless,  in the past   ratios $\frac{M_{\mathrm{std}}}{M_{SIV}}$ 
   as high as 80   \citep{Karachantsev66} or  mass-luminosity ratios $M/L$
   of about 700 \citep{Bahcall74} have been obtained.

The conclusion of this analytical study is that scale invariant dynamics in our low density Universe could  account
 for the high observed velocities of galaxies in clusters, despite the reduction factor $\psi(\tau)$ in the equation of motion. 
Of course, numerical simulations would be needed  to further 
quantify the scale invariant effects; at least the present analytical developments are an encouragement to the undertaking 
of such calculations. Globally, it appears in this new context that the hypothesis of  large amounts of dark matter
may  no longer be necessary, the dynamical acceleration $\left(\frac{\psi(\tau)}{\tau_0}   \frac{d\bf{r}}{d\tau} \right)$
producing a sufficient increase of the velocity dispersion during the ages.   Now, we will examine whether such a  conclusion is supported by  
other dynamical problems.

\section{Galactic rotation}

\subsection{The possible evolution of the rotation curves: theory and observations}   \label{rotA}

In the outer regions of spiral galaxies, the orbital motions should  in principle be  close to Keplerian behaving like $1/\sqrt{r}$
being dominated by the central mass concentration. The large differences from what is  generally observed with the 
flat or relatively flat  rotation curves over a few tens of kpc (see review by  \cite{Sofue01}) have greatly promoted, since the 
beginning, the hypothesis
of dark matter \cite{Trimble87}. Such dark components  have been searched  over the last forty years. 
We have already considered the topic previously \cite{Maeder17c}, but new 
observational and theoretical results have considerably 
enriched and strengthened the subject.

\subsection{Properties of orbital motions} \label{twobody}

The  two-body problem  in SIV theory has been studied by \cite{MBouvier79} 
and \cite{Maeder17c}.
The equation of motion (\ref{Nvec}) becomes in 
 plane polar coordinates $r$ and $\varphi$,
 \begin{eqnarray}
 \ddot{r} - r  \dot{\varphi}^2  = - \frac{G \, M}{r^2} +\kappa(t)  \dot{r} , \quad \quad 
 r \ \ddot{\varphi} + 2 \ \dot{r}  \dot{\varphi}  =  \kappa(t)  r  \dot{\varphi}.
 \label{Ntheta}
 \end{eqnarray}
The time variable $t$ is provisionally considered, since the equations are much simpler; however we turn to the 
current time units $\tau$ for the main conclusions below.
 Integration  gives the equivalent law of angular momentum conservation,
\begin{equation}
 \kappa \, r^2\, \dot{\varphi} = L=const. \quad \mathrm{with} \; \;  \kappa=1/t.
\label{consang}
\end{equation}
 Some manipulations give
 the corresponding Binet equation \cite{MBouvier79},
\begin{equation}
 \frac{d^{2} u}{d \varphi^{2}}  + u \, = \, \frac{GM \kappa^{2}}{L^{2}} \,, \quad \mathrm{with} \; \; u= 1/r .
 \label{Binet}
 \end{equation}
 The second member is scaling like $\lambda \sim 1/t$. This equation as in the classical case
 is leading to a family of conics,
\begin{equation}
  r(\varphi) = \frac{p}{1+e\; cos \varphi}, \; \; \mathrm{with} \; p =\frac{L^{2}}{GM \kappa^{2}}= \frac{b^2}{a},  \;   b=  a \sqrt{1-e^2}.
\end{equation}
 $a$ and $b$ are  the semi-major and semi-minor axes. Eccentricity $e$ is scale invariant;
   for $e=0$, one has  $r =p$.   Both 
$p$ and   $r$ are of power -1 and thus slightly increase with time. Differentiating $r$ gives,
\begin{equation}
\frac{\dot{r}}{r} \, = \, 1/t \, \quad \mathrm{implying} \;
r \, \sim \, t \, ,  \; \mathrm{and} \; \frac{\Delta r}{r} \, = \, \frac{\Delta t}{t} \,.
\label{tr}
\end{equation}
The  orbital motion  of a bound system is  a circle or an ellipse
with a  slight outwards spiraling  motion. With the above  $t - \tau$ relations (\ref{T2}) and (\ref{dT1}) 
we get in current units,
 \begin{eqnarray} 
\frac{dr}{r}  = \frac{dt}{t}  =  \frac{dt}{d\tau} \frac{d\tau}{t} =
\ \frac{(t_0-t_{\mathrm{in}})}{(t_{\mathrm{in}} + \frac{\tau}{\tau_0} (t_0- t_{\mathrm{in}}))}
\frac{d\tau}{\tau_0} = 
  \psi(\tau)\frac{d\tau\,}{\tau_0} .
\label{dr}
\end{eqnarray}
\noindent
where $\psi(\tau)$ was already defined above (\ref{pss}).
At  present time $\tau_0$, the relative change of an orbital radius is
\begin{equation}
\frac{\Delta r}{\Delta \tau} = \psi_0 \, \frac{ r_0}{\tau_0}, \quad \mathrm{for\;a\;  circle \;  or\; an\; ellipse.}
\label{dr4}
\end{equation}
 At a  time $\tau$, one has $\psi(\tau)$ instead of $\psi_0$.
The relation between $r$ and $\tau$ is not linear, due to the term $\psi(\tau)$. However, up to time intervals of 400 Myr  the deviations from 
$\psi_0$ are of the order of 1\%, see also Eq.~(\ref{ratiogal}).
 
 Remarkably,  the effect of the additional acceleration term  just compensates
 the  usual slowing down due to orbital expansion, thus keeping the orbital $\upsilon$ constant,
\begin{equation}
\upsilon = r \,  \dot{\varphi}=  \frac{ L}{r \, \kappa} = \frac{GM}{r}= const. \quad \mathrm{since} \; \;
 r \sim t \; \mathrm{and} \; \kappa=\frac{1}{t},
\label{ups}
\end{equation}
\noindent
an effect already found by \cite{MBouvier79}.

\subsection{Application to galactic rotation}  \label{galrot}

{\bfseries{Fig.~\ref{MW} shows several observed rotation curves of the Milky Way as a function of the galactocentric 
distance $R(kpc)$, in red the results by Huang et al. \cite{Huang16}.}} Up to about 26 kpcs  there are some 
dips and bumps in the rotation curve, likely corresponding to spiral arms. 
Beyond 26 kpc a progressive decrease is observed
up to  distances as large as 100 kpc. This extended  rotation curve 
  is based on the observations of about 16'000 red clump giants in the outer disk and 5700 K giants in the halo.
{\bfseries{A proper account of the anisotropic motions is performed in the analysis, which also permits to see the undulations due to spiral arms.
 Fig.~\ref{MW} also  shows the more recent results by Eilers et al. \cite{Eilers19} and by Jiao et al. \cite{Jiao23},
who both find steeper curves for the 5 to 25 kpc inner distance interval. 
The study  by Eilers is based on the 6D data (location and velocities) of more than 23'000 red giant stars
where the velocity curve is   determined from Jeans equation for an axisymmetric  gravitational potential. The authors found a decreasing velocity of 
1.7 $(\pm  0.1)$ km/s. They estimated   a value of 7.25 ($\pm 0.26)\cdot 10^{11}$ M$_{\odot}$ for 
 the total mass of the Milky Way, and that this mass is still dominated by dark matter
for radii larger than about 14 kpc. The results by Jiao et al. are based on the data from Gaia (Gaia DR3). They find  a steeper decrease of the
velocity  of 30 km s$^{-1}$ between 19.5 and 26.5 kpc as illustrated in the 
figure; this suggests an absence of significant mass increase at radii larger
than 19 kpc. This also leads to a smaller mass value for the Galaxy
of 2.06$^{(+0.24)}_{(-0.13)} \cdot 10^{11}$ M$_{\odot}$. It is interesting to notice that up to 20 kpc from the center,
the data by Eilers et al. and by Jiao et al.  agree with the lowest points of th undulations due to galactic arms by Huang et al., while the
undulations are absent from the other two studies.
}}

 Some basic properties of the scale invariant  two-body mechanics have been recalled in Sect. \ref{twobody}.
With these laws we can  reconstruct the evolution of the galactic rotation over the ages due to
scale invariant effects:
\begin{itemize}
\item The orbital  radius 
(here the galactocentric radius)  increases with time  (Eq.~\ref{dr}). 
\item The orbital velocity $\upsilon$ of a given star 
around the galactic center is keeping constant during the additional expansion  (Eq.~\ref{ups}). 
\end{itemize} 
If the time intervals considered in the evolution of galaxies are not negligible with respect
to the age of the Universe, Eq.~(\ref{dr4})  cannot be used;  for such cases  we must integrate 
Eq.~(\ref{dr}) over time $\tau$ between $\tau_1$ (in Gyr)
the epoch considered in the past with size $r_1$, and the present time $\tau_0$ with size $r_0$,
\begin{eqnarray}
\int^{r_0}_{r_1} \frac{dr}{r} = \frac{(1-\Omega_ {\mathrm{m}})^{1/3}}{\tau_0} \int^{\tau_0}_{\tau_1} 
\frac{d\tau}{\Omega_{\mathrm{m}}^{1/3}+\frac{\tau}{\tau_0}(1-\Omega_ {\mathrm{m}}^{1/3})}=  
\int^{\tau_0}_{\tau_1} \frac{d\tau}{\tau_0  \frac{\Omega_ {\mathrm{m}}^{1/3}}{(1- \Omega_ {\mathrm{m}})^{1/3}}+ \tau}=
\int^{\tau_0}_{\tau_1}\frac{d\tau}{\tau_{\Omega}+\tau},
\end{eqnarray}
where we use the constant $\tau_{\Omega}$ defined in Eq.~(\ref{Nvecv}). The integration gives
\begin{equation}
\quad \quad \quad  \quad   \quad \quad \quad \frac{r_0}{r_1} \,=\,  \frac{\tau_{\Omega}+\tau_0}{ \tau_{\Omega}+\tau_1}.
\label{ratiogal}
\end{equation}
For example, for  $\Omega_ {\mathrm{m}}=0.045 $ \cite{Planck20}, $\tau_0=13.8$ Gyr  and   an age  $\tau_1$ of 6 Gyr, 
 one has $\tau_{\Omega}= 7.62$ Gyr and a scaling of the distances $\frac{r_0}{r_1}= 1.57$. This means that at an age of 6 Gyr, the same velocity 
as one on the red curve was reached at a galactocentric distance $r_1$ a factor 1.57 smaller than the present distance $r_0$.
This enables  the construction of the past rotation curves of galaxies at different past ages 0.2, 3, 6, 9 Gyr, as shown
by the blue curves in Fig.~\ref{MW}, on the basis of the present
red curve based on data by \cite{Huang16}, (the only one we have at large distances).
Among the blue curves, the left one corresponds
to the early age of galaxy formation (0.2 Gyr). {\bfseries{We are not making  here a critical analysis of the different curves, which is beyond the scope
of the paper, but just using the longest of the available recent  curves to illustrate the cosmic evolution resulting from scale invariant effects.  \\

\begin{figure*}[t!]
\centering
\includegraphics[width=12cm, height=8cm]{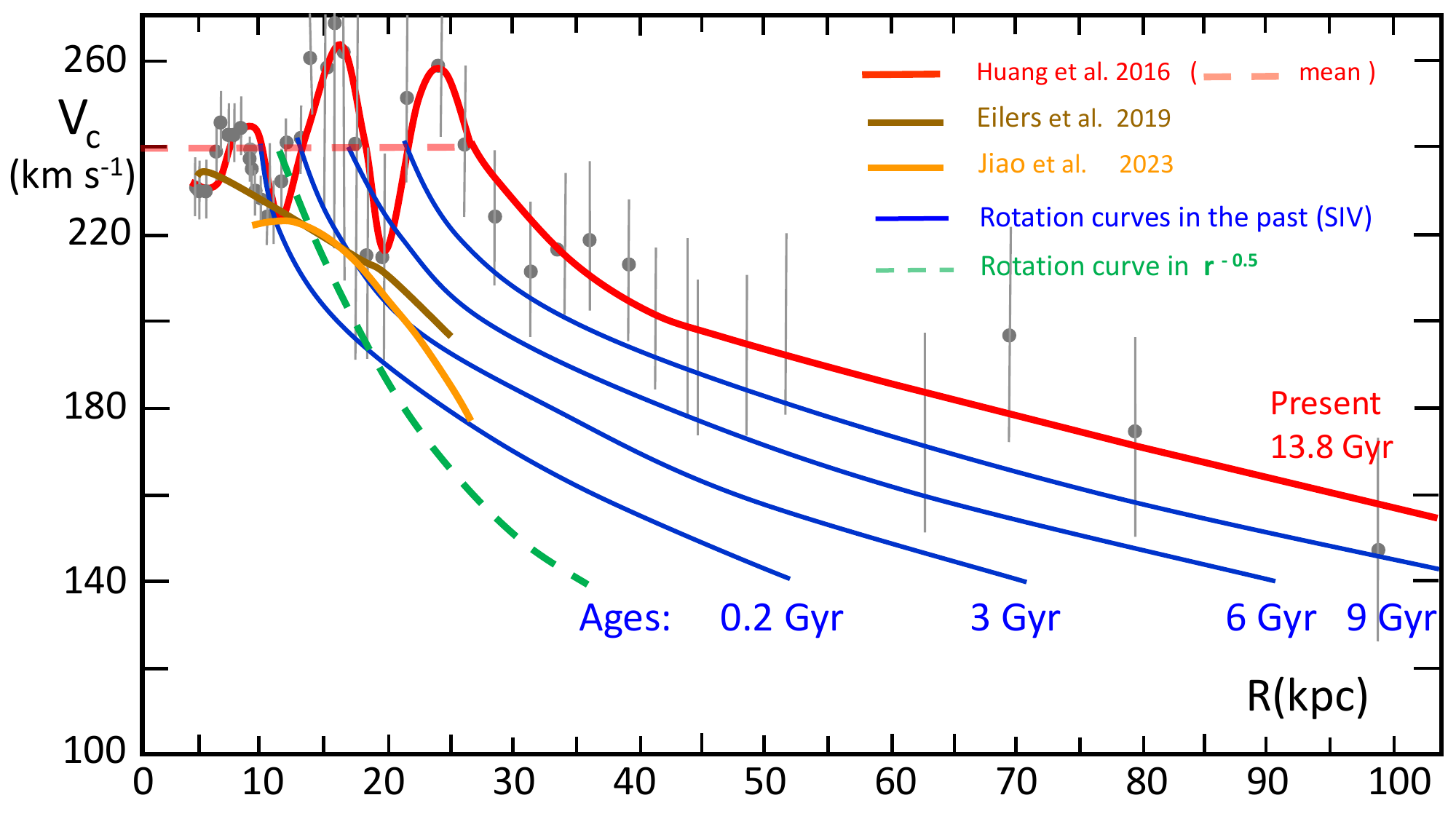}
\caption{Evolution of the rotation curve of the Milky Way. The grey points are the observed velocity
 averages by Huang et al. \cite{Huang16} with their error bars.  The thick
red line represents a mean rotation curve. The thin broken red line describes the flat mean of the undulations   of the  velocity distribution
up to a radius of 26 kpc. The  brown and orange lines show respectively the results
of Eilers et al. \cite{Eilers19} et of Jiao e al. \cite{Jiao23} for the inner galaxy.
The blue  lines show
 the  rotation curves  predicted by   the scale invariant theory for  different ages in the past history of the Universe, starting backwards from the red curve.
 Calculations have been performed with no dark matter in a  model with  $\Omega_{\mathrm{m}}=  \Omega_{\mathrm{b}}$, where $\Omega_{\mathrm{b}}=0.045$.
 The thick broken green line shows a Keplerian curve  in  $1/\sqrt{r}$ near  the  age of galaxy formation.}
\label{MW}
\end{figure*}

These results show that the effects of scale invariant evolution with increasing radii at constant 
orbital velocity  are favouring the flatness of the rotation curves. 
Conversely, when going back in time these  curves  become progressively  steeper  and steeper
for smaller ages.}}
We can draw the following predictions about the past effects of scale invariance in galaxies  from the above Fig.~\ref{MW}:
\begin{enumerate}
\item In the past, the  extensions of galaxies  were smaller than today. Galaxies were more compact
 according to the ratio (\ref{ratiogal}).

\item Some part of the  flat rotation curves  is resulting  from the cosmic expansion
associated to scale invariance.

\item In the past, the rotation curves started to decrease at much smaller distance from the galactic center than today, 
 with a  much steeper profile, {\bfseries{suggesting }} smaller fractions of dark matter in the past.
\end{enumerate}
This describes only the effects resulting from scale invariance and does not preclude all the other current effects of stellar dynamics
to operate over the ages.   \\

It is interesting to compare these theoretical results with some recent observations.
 Six star forming galaxies with redshifts between  $z=0.8 - 2.4$ were analyzed  by \cite{Genzel17}, 
 and a large sample of 101 other galaxies from the KMOS and SINS/zC-SINF surveys with redshifts
  between $z=0.6$ and 2.6 were studied by \cite{Lang17}. Both surveys give trends corresponding  to the above 
 point 3: the observed past rotation curves   were  not flat, but   were exhibiting 
  steeply decreasing  velocities with their galactocentric radius. Thus,
 even at distances of several  effective radii, the authors 
  found that the dark matter fractions were modest or negligible in the past. 
  A variety of discussions and interpretations of the above results  
  have taken  place (see next ref.). However,
 a further analysis by \cite{Genzel20} from the same survey confirmed that 
  at $z \sim 2$ galaxies have very low dark matter fractions ($< 20$\%).

 The observations  are also improving rapidly, and  very recently the above results
  were confirmed by new data from  different instruments and wavelengths.  In particular,
  a  hundred galaxies at the peak of their star formation ($z= 0.6 - 2.5$) were observed with
    ESO-VLT, LUCI, NOEMA at IRAM and ALMA, and further analyzed
   so doubling  the previous samples \cite{Nestor Shachar23}. 
  These new  observations show rather steep rotation curves, which according to the authors
 demonstrate that the fractions of dark matter strongly decrease 
  with increasing redshifts: for example
  at $z = 0.85$  the  fraction of dark matter is estimated to be 38\% of the typical value,
    and this fraction goes down to 17\% at $z= 2.44$.


  In standard models with dark matter, the above decrease  of DM with redshifts is difficult to interpret.
  It is not clear where at high $z$   the usual dark matter fraction is located, in order to let only such small
  dynamical signatures. The possibility that the missing dark matter is lying in more external
  regions would be  a rather  troublesome one. Indeed, in the CDM theory  the baryons set in
  the potential wells already created by dark matter during the radiative era.  Thus,  it is not clear why 
galaxies would contain less dark matter in the past, since the accumulation of DM is considered preceding that of baryons.

    At the opposite, we point out
   that the above mentioned results are    corresponding to the predictions of 
  the SIV theory concerning the evolution of rotation curves (Fig.~\ref{MW}).
  In itself  it is not at all a proof of the theory, also because the variety of different results concerning the rotation curve, but  an interesting indication,
 which is  perfectly consistent with previous results on the
  origin of the high velocities of galaxies in clusters (Sect. \ref{clust}),
and with other results quoted in the introduction, in particular the growth of density fluctuations \cite{MaedGueor19}.

  \subsection{The limited connection with MOND} \label{mondrot}
   
  The modified Newton equation, in the approximation of a constant scale factor $\lambda$,
   is leading to the deep MOND limit  \cite{Milgrom09} and \cite{Milgrom14} with a gravity given by \citep{Maeder23},
  \begin{eqnarray}
  g = \sqrt{a_0 \; g_{\mathrm{N}}},  \quad \mathrm{and} \; \; \;
   a_0 = \frac{n c \,H_0 \, (1-\Omega_{\mathrm{m}})^2}{4}, \quad  \label{mondo} \\
\mathrm{with} \; H_0=\frac{2}{\tau_0} \frac{ (1-\Omega^{1/3}_{\mathrm{m}})}{(1-\Omega_{\mathrm{m}}) } \quad 
  a_0 = \frac{nc \, (1-\Omega_{\mathrm{m}}) (1-\Omega^{1/3}_{\mathrm{m}})}{2\, \tau_0},
  \label{mond}
  \end{eqnarray}
where $g_{\mathrm{N}}$ is the Newtonian gravity.
  There $n$ is of the order of unity and $c$ is  the velocity of light.
  As  the  range of $\lambda$-variations is  limited (Sect. \ref{neuton}),
   over  timescales  less than about  $2 \times 10^8$ yr the scale factor $\lambda$ undergoes changes smaller than 1\%. 
For example, in Universe models with a value  $\Omega_{\mathrm{m}} =0.3$, 
  for galaxies with dynamical timescales  ({\emph{e.g.}} orbital period)
  of 200 Myr,  the approximation of constant $\lambda$ is in error
   of only 0.5 \%. For the extreme case with $\Omega_{\mathrm{m}} =0.045$, the corresponding errors amount to 1\%. 
  Thus, MOND may be considered as an    approximation tangent to the SIV  theory over one to two  galactic rotations.
However, within SIV the  approximation is not valid over longer timescales covering the global evolution of a galaxy.
     
In SIV theory, the deep limit regime is entered when the dynamical acceleration (the additional term in the modified Newton equation~(\ref{Nvec4})) becomes larger
   than the usual Newtonian gravity term \citep{Maeder23}, (this limit may also be influenced by the opposite gravity of nearby objects). 
This condition leads to the above value of $a_0$  (Eq.~\ref{mondo}). The product $c \, H_0$ is equal to 6.80 $10^{-8}$ cm s$^{-2}$ 
for $H_0= 70 $  km/(s Mpc).
 For $\Omega_{\mathrm{m}}$=0, 0.10, 0.20, 0.30 and 0.50, with $n \approx 1$
   one has $a_0 \approx$ (1.70, 1.38, 1.09, 0.83, 0.43) $\; \times \;  10^{-8}$ cm s$^{-2}$ respectively.
 These well encompass the observed value
   $a_0$ of about 1.2 $\times \,10^{-8}$ cm s$^{-2}$ \cite{Milgrom14}.

There are some similarities, but also differences, between the two theories. In MOND,  equation $g = \sqrt{a_0 \; g_{\mathrm{N}}}$  directly leads to a constant
circular velocity $\upsilon^2 = \sqrt{a_0 GM}$
 independent from the orbital radius in the low gravity regime.  In SIV theory, the two-body system predicts a constant orbital velocity 
during the increase of radius resulting from the expansion due to the scale invariant effects  (\ref{ups}).
This is not the same:  in SIV  the velocities may  depend on their initial values and on the duration of the expansion, i.e. 
the ages.   Moreover, in SIV it is not necessary that the gravity is smaller than $a_0$ to have   deviations from Newtonian.
 The dynamics is always governed by the full equation~(\ref{Nvec4}) with both Newtonian and dynamical accelerations.

Turning to observations, we see that  the galactic rotation curve (Fig.~\ref{MW})  indicates that the flat part is relatively limited. 
Over large  distances,  the orbital velocity is not a constant,  but  a slowly decreasing function.
Also MOND is not in agreement with  the new results by the Genzel group presented above.
The deep limit, which accounts for the flat 
rotation curves, applies in principle the same way at all epochs, while the observations of galaxies at $z=2$ do not show flat curves
thus predicting much less dark matter.
The  correspondence between MOND  and SIV   appears  limited to objects of low redshifts and to  a certain domain of radii in galaxies
  (a few tens of kpcs).

   In the case of the  clusters of galaxies,  the MOND-like approximation should  barely 
  apply, since their dynamical timescale is much longer than for galaxies due to their lower densities. 
   For Universe models with a value  $\Omega_{\mathrm{m}} =0.3$, 
  clusters with dynamical timescales  of 1, 3 or 5 Gyr,  
  the approximation of constant $\lambda$ would be  in error
   of 2.4, 7.1 and  11.9 \%. For the case with 
   $\Omega_{\mathrm{m}} =0.045$, 
   the corresponding errors amount to 4.6, 14.0 and 23.3 \% respectively. 
   Thus, MOND might   be a possible approximation 
   of scale invariant   dynamics for small clusters, but not for large clusters,
   with  dynamical timescales of the order of the age of the Universe.

\subsection{The connection between the assumed dark matter and the baryons:
the signature of a gravity effect}  \label{dmbar}

  An  interesting observation
 concerns the radial acceleration  (RAR) studied by Lelli et al. \cite{Lelli17}, who showed that 
  the relation between average accelerations $\frac{\upsilon^2}{R}$  and $\frac{GM}{R^2}$ 
  for galaxies deviates from a 1:1 relation for gravities below $a_0$.
  The RAR  is well
  accounted for  by both MOND (cf. Lelli et al.) and the  SIV theory \cite{MaedGueor20b}, with however
  a possible difference concerning the extreme case of spheroidal galaxies. 
Remarkably as found by
  Lelli et al., see  also \cite{Chan22}, the amount of dark matter (DM, measured by the deviations from the 1:1 line) seems to be always
  directly determined by the amount of baryonic matter, although in total 2700 measurements
  have been made in very different galaxies and {\emph{at different distances from their centers}}.
  Such a result is  pointing in favour of a gravity effect rather than for DM.

As mentioned above, the same kind of relation between the  dynamical
and  baryonic masses of central regions is also present in clusters of galaxies \cite{Chan22}. 
Moreover, such a relation is also observed in redshifted galaxies, 
where Nestor Shachar et al. \cite{Nestor Shachar23} are finding  a strong correlation 
between the fraction of DM and the baryon surface density within the 
effective radius (the part corresponding to the flat curve). The convergence 
of such  findings in  three  so different environments confirms the intimate   relation between the baryons
and the assumed dark matter. It is clearly supporting  the views that this general connection is the signature of gravity effect.

 \section{Dispersion of the vertical velocities}

The velocities are too high not only in the galactic plane as seen above, but also in the  "vertical"  direction of the W-motions perpendicular
to the galactic plane. 
The old problem of the dispersion of vertical velocities  
 cannot be considered as a test,
since its 70 year history has shown that a lot  of  astrophysical   effects may  significantly intervene,
as recalled in the introduction.  Scale invariant effects, if there would be
 no interaction with other effects, only account for a small fraction
of the total vertical velocity dispersion, thus confirming the richness of the involved processes.

Spitzer and Schwarzschild \cite{Spitzer51}, who showed that the  "vertical" velocity dispersion (perpendicular to the
galactic plane) was  increasing with the age of the stellar group  interpreted this effect as 
 due to collisions with large interstellar clouds.
 This effect was  confirmed by  further  studies, see Seabroke \& Gilmore \cite{Seabroke07}. Among the  other
astrophysical effects  proposed to account for the growth of velocities with age, we may mention 
the effects of spiral waves by Barbanis \& Woltjer \citep{Barbanis67},
 of the driving be an unknown diffusive process by Wielen \cite{Wielen77},
of massive halo black  holes by Lacey \& Ostriker \cite{Lacey85},
 of mergers of dwarf galaxies by Toth \&  Ostriker \cite{Toth92}, of evaporating star clusters by Kroupa \cite{Kroupa02} and
of the evolution of the interstellar medium by Kumamoto et al. \cite{Kumamoto17}.

\begin{figure}
\centering
\includegraphics[angle=0.0,height=9.5cm,width=12cm]{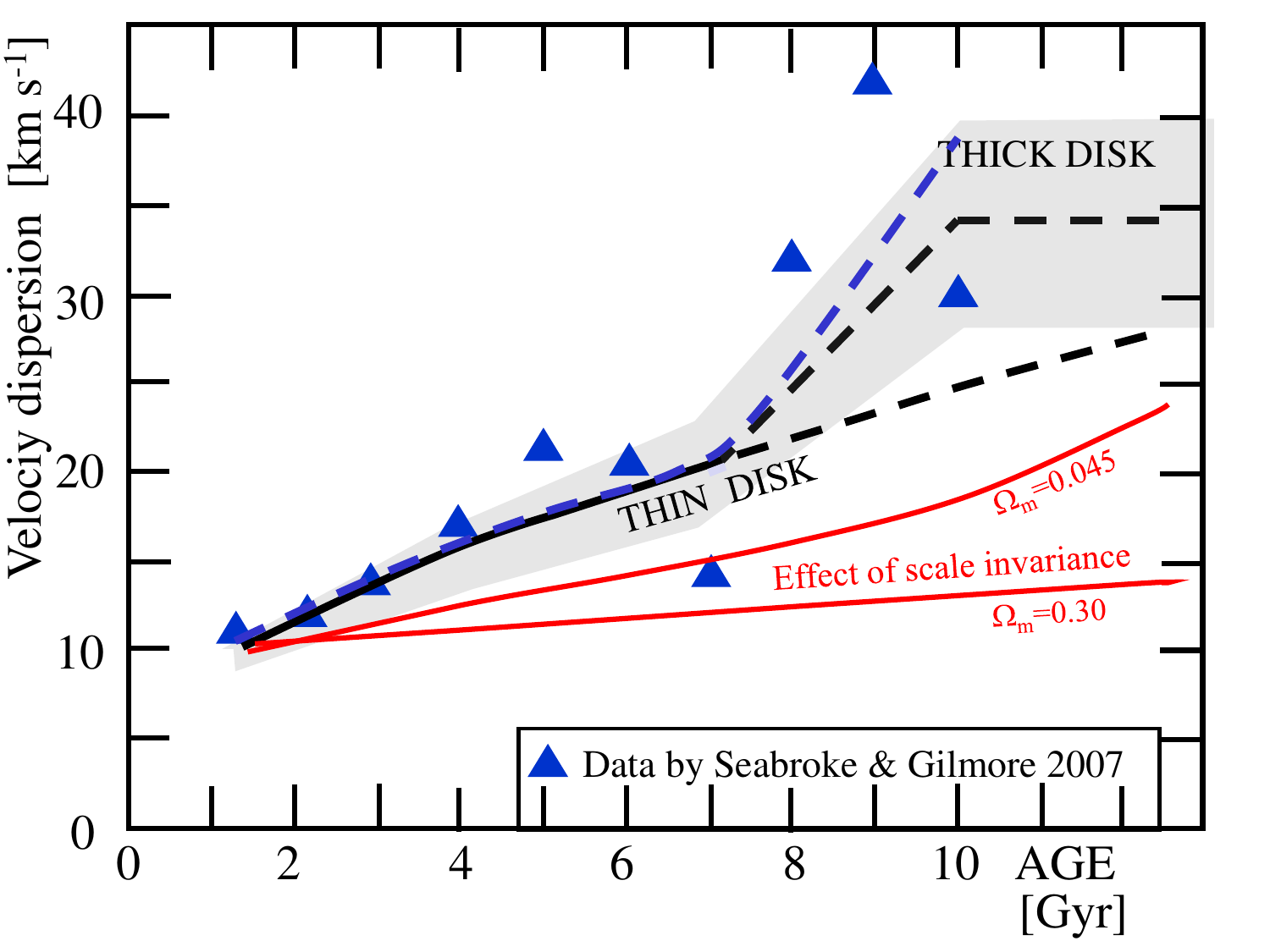}
\caption{The vertical velocity dispersion as a function of the age of the star population considered.
The thin and the thick disks are represented by  black broken lines according to the  WD data by \cite{Cheng19}, see also
\cite{Raddi22}. Data from \cite{Seabroke07} are shown as blue triangles (with a broken blue line
for their average trend),  these are in  good agreement
with the WD data for  the thin disk.  
 The results for scale invariant  effects are represented 
by   red lines for $\Omega_{\mathrm{m}}= 0.30$ and 0.045 in absence of dark matter;
as usual, the effects are smaller for denser Universe models.}
\label{vdisp}
\end{figure}

  Recent valuable observations of the age effect on the  dispersion of vertical 
  velocities have  been providedby Cheng et al. 
  \cite{Cheng19} and  by Raddi et al. \cite{Raddi22}  on the basis of the kinematical data 
  for white dwarfs from Gaia DR2 in the solar neighborhood. 
  Their  data are in general agreement with the results of \cite{Seabroke07} based on the large survey of nearby F and G dwarf stars
by  \cite{Nordstrom04}. In Fig.~\ref{vdisp}, the agreement is excellent for the thin disk up to 
  ages of about 7 Gyr. For larger ages, the scatter of the data is larger,  but both samples well illustrate 
the large increase of the vertical velocity dispersion for the old star population of the thick disk.
As mentioned in the introduction, a number of authors have identified  many 
astrophysical effects which may possibly contribute to the increase with time 
of the observed vertical velocity dispersion.

Some  characteristics of the acting astrophysical processes have been enlightened by \cite{Seabroke07}.
  They pointed out that the effect \cite{Spitzer51}
 of collisions with giant clouds saturates 
  after some time with no further increase of the dispersion. 
  They also  demonstrated that the effect producing the growth of the dispersion 
   is acting continuously  throughout the motions 
  around the galactic plane  and not only when the stars are crossing it.

  A study of the scale invariant effects in the oscillations of stars around the galactic plane  has been performed by \cite{Magnenat78}. 
From the modified Newton equation (\ref{Nvec}),  they showed that
  the vertical motions in the $z$-directions obey the following equation of motion  (in variable $t$),
  \begin{equation}
   \ddot{z}  - \frac{1}{t}  \dot{z} + \omega^2(t)  z = 0, \quad
   \mathrm{with} \; \omega^2(t) = \left(\frac{\partial K_z}{\partial z}\right) = 4 \pi \, G \varrho \, .
   \label{equz}
   \end{equation}
   There, 
   $K_z$ is the vertical component of the acceleration, $\varrho$ 
   the mean spatial density at the level of the galactic plane and $\omega$ the oscillation frequency.
  The oscillatory solutions around the galactic plane are of the form,
   \begin{equation}
   z = \left(\frac{z_{\mathrm{form}}}{t_{\mathrm{form}}}\right) \, t \, \sin (s \, \ln t+\varphi), 
   \quad \mathrm{with} \quad  s =  \sqrt{\omega^2_0 \, t^2_0-1} \, ,
   \label{z}
   \end{equation}
   where the index ``form'' indicates the time of formation and ``0''  the present time.
   This shows that the oscillation amplitudes are increasing with time t.
   The amplitudes of the velocity when crossing the galactic plane behave like,
   \begin{equation}
  W \equiv  \dot{z} \sim  \frac{z_{\mathrm{form}}}{t_{\mathrm{form}}} \,
   \sim \frac{1}{t_{\mathrm{form}}}, \quad \mathrm{thus} \;  
   W(t_{\mathrm{form}})= W(t_0) \frac{t_0}{t_{\mathrm{form}}},
\label{w}
\end{equation}
under the assumption  that stars preferentially form in the galactic plane. If the stars are formed outside the galactic plane, 
i.e. at a larger $z_{\mathrm{form}}$, the velocity is increased as discussed below for the thick disk. There,
 $ W(t_{\mathrm{form}})$  is the vertical velocity (in the galactic plane)  nowadays
 of a star born at time $t_{\mathrm{form}}$, while $W(t_0)$ is the velocity of
 stars recently born. Turning now to the current time units with (\ref{T2}) , we get  the velocity $ W(\tau_{\mathrm{form}})$,
 \begin{equation} 
   W(\tau_{\mathrm{form}})= W(\tau_0) \, \frac{1}{t_{\mathrm{in}}+
   \frac{\tau_{\mathrm{form}}}{\tau_0} (1- t_{\mathrm{in}})}.
\label{wo}
\end{equation}
This possible cosmological effects of scale invariance are illustrated 
 by the  red lines in Fig.~\ref{vdisp}, for the cases $\Omega_{\mathrm{m}}=0.045 $ and 0.30.  
The account in variable $\tau$ of the limited variations
of $\lambda$  reduces the effect,  compared to the results by \cite{Maeder17c}.
  As the observations show a steep initial increase from about  10 km/s
during the first Gyr, we start with a value of 10 km/s at an age of 1 Gyr.
Fig.~\ref{vdisp} shows a small regular increase  
 of the dispersion: for $\Omega_{\mathrm{m}}=0.045$
  it is  of 4.9 km/s at an age of 7 Gyr, at the end of thick disk phase.
   For $\Omega_{\mathrm{m}}=0.30$, the increase amounts to 2.0 km/s,
 which  only accounts for a fraction of the observed effect.
  
  Clearly, this lets room for 
 other effects. The thickness of the disk at the time of star formation is a key point.
 This is shown by  the value  $z_{\mathrm{form}}$ 
 in the first of Eqs.~(\ref{w}): if the stars are ``launched''
 at their birth from a more distant place, the amplitudes will be proportionally larger. This accounts
 for the thick disk effect in Fig.~\ref{vdisp}.
 Also, the collision with massive interstellar clouds, partially favoring 
 an equipartition of the energy, would speed
 up the velocity by about 7 km/s at an age  of 7 Gyr according to \cite{Spitzer51}.
 The other effects mentioned
 above would also contribute to several km/s at 7 Gyr,
  in particular the formation of star clusters \cite{Kroupa02}.
 The gravitational field associated to the spiral structure would
  also  contribute \cite{Barbanis67}, but probably  more significantly for the
   horizontal dispersion of the  $U$ and $V$ components.
   
   The conclusion we get from Fig.~(\ref{vdisp}) is that scale invariant 
     may  play some role. However, this role is for now probably largely underestimated since the red lines show these effects 
   without any  interaction with the other dynamical mechanisms and this is a major point.
   Indeed, the additional acceleration term in the modified Newton equation~(\ref{Nvec4})
   depends on the effective total velocity, thus the influence of  scale invariance would be magnified
   all the way through the ages by its interaction with the  various  dynamical  processes.
    The solution of these interactions is beyond the scope
   of the present study, where only the  effects acting on an initial velocity 
   of 10 km/s, typical of the situation after 1 Gyr, were considered.
   Nevertheless, it shows that, if scale invariant effects do exist,
   they are significant actors of the problem.

\section{Conclusion}

Overall, there is an impressive convergence of results when interpreting well-known astrophysical facts: the observed excesses of velocity in clusters of galaxies, the observed flat rotation curves of low redshift galaxies, the observed steeper rotation curves of high-redshift galaxies, the unicity of the relation between dark and baryonic matters repeatedly observed in galaxies  \cite{Lelli17}, in galaxies and galaxy clusters \cite{Chan22} and in significantly redshifted galaxies  \cite{Nestor Shachar23}.  All these results come from a broad range of sources and authors,  and happen in very different astronomical objects spanning themselves a vast range in stages of evolution, masses and spatial scales. It would be estonishing that Nature "conspires" to make all these results pointing to the same direction for no reason. This is in fact questioning  the very existence of dark matter.

 These results correspond to the predictions
 of the scale invariant theory, which is resting on a  simple fundamental principle: 
enlarging the group of invariances subtending the theory of gravitation as suggested by Dirac \cite{Dirac73}.
Moreover, the  theory contains no "adjustment parameter".
The resulting  effects depend on the inverse of the age of the Universe and are thus negligible
in current life, while their cumulative effects may play some role in  the  cosmic evolution. They are also consistent with the results of some other tests  recalled in the introduction. The global conclusion is that it is  worth  pursuing further observational tests of scale invariance and to perform numerical simulations.
 It would be a pity to miss the possibility offered by this additional symmetry over standard $\Lambda$CDM.\\

Acknowledgements:

      The author expresses his gratitude to Prof. James Lequeux for his support and encouragements, to Dr. Vesselin Gueorguiev 
for fruitful collaboration over the years and to Prof. Frederic Courbin for very helpful comments and continuous support.



\begin{thebibliography}{99}




\bibitem{Bahcall74} Bahcall, N., \emph{The Perseus Cluster: Galaxy Distribution, Anisotropy, and the Mass/Luminosity Ratio}, ApJ, 187, 439 (1974)

 \bibitem{Barbanis67} Barbanis, B., Woltjer, L. , 
\emph{Unresolved stellar companions with Gaia DR2 astrometry}, 
MNRAS, 496, 1922 (2020)


  \bibitem{Blindert04} Blindert, K., Yee, H.K.C., 
  Gladders, M.D. et al., {\emph{Dynamical masses of RCS galaxy clusters}}, IAU Coll. 195,  p. 215  (2004)


\bibitem{BouvierM78} Bouvier, P., Maeder, A., \emph{Consistency of Weyl's Geometry as a Framework for Gravitation},
 Astrophys.  Space Science, 54, 497 (1978)



\bibitem{Canuto77}
{Canuto}, V., {Adams}, P.~J., {Hsieh}, S.-H., \& {Tsiang}, E.,
\emph{Scale-covariant theory of gravitation and astrophysical applications}, Physical Rev. D, 16, 1643 (1977)

\bibitem{Carroll92}
{Carroll}, S.~M., {Press}, W.~H., \& {Turner}, E.~L., \emph{The cosmological constant}, Annual Rev. Astron. Astrophs., 30, 499 (1992)

\bibitem{Chan22} Chan, M. H., \emph{Two mysterious universal dark matter-baryon relations in galaxies and galaxy clusters},
 Physics of the Dark Universe, 38, 101142 (2022)

\bibitem{Cheng19} Cheng, S., Cummings, J.D., Menard, B., \emph{A Cooling Anomaly of High-mass White Dwarfs}, Astrophys. J. 886, 100 (2019)

\bibitem{Chiu16}Chiu, I., Mohr, C.L., McDonald, M.  et al., \emph{Baryon content of massive galaxy clusters at 0.57 < z < 1.33}, MNRAS, 455, 258
(2016)



\bibitem{Dirac73}
Dirac, P.~A.~M., \emph{Long range forces and broken symmetries}, Proceedings of the Royal Society of London Series A,
333, 403 (1973)

\bibitem{Dirac74}
Dirac, P.~A.~M., \emph{Cosmological Models and the Large Numbers Hypothesis},
 Proceedings of the Royal Society of London Series A, 338, 439
(1974)

\bibitem{Eddington23} Eddington, A.S., \emph{The Mathematical Theory of Relativity}, Chelsea Publ. Co. New York, 266 p. (1923)

\bibitem{Eilers19} Eilers, A.-C., Hogg, D.W., Ris, H.W. et al., \emph{The circular  velocity curve of the Milky Way from 5 to 25 kpc}, Astrophys. J. 871, 120 (2019)

\bibitem{Einstein18} Einstein, A., \emph{Kommentar zu "Hermann Weyl: Gravitation und Elektrizit\"{a}t"},
Sitzung Berichte der Königlich Preussischen Akademie
des Wissenschaften (Berlin), p. 478. (1918)

\bibitem{Famaey12} Famaey, B., McGaugh, S.S., \emph{Modified Newtonian Dynamics (MOND): Observational Phenomenology and Relativistic Extensions}, Living Rev. Relativity, 15, 10 (2012)







\bibitem{Frieman08}
Frieman, J.~A., Turner, M.~S., \& Huterer, D., \emph{Dark energy and the accelerating universe}, Annual Rev. Astron. Astrophs.,  46, 385 (2008)



\bibitem{Ge16} Ge, C., Wang, Q.D., Tripp, T.M. et al., \emph{Baryon content and dynamic state of galaxy clusters: XMM-Newton observations of A1095 and A1926}, MNRAS, 459, 366 (2016)


\bibitem{Genzel17} Genzel, R., Forster
Schreiber, N.M., Ubler, H., \emph{Strongly baryon-dominated disk galaxies at the peak of galaxy formation ten billion years ago}, Nature, 543, 397
(2017)

\bibitem{Genzel20} Genzel, R., Price, S. H., Ubler, H., et al. \emph{Rotation Curves in z  1-2 Star-forming Disks: Evidence for Cored Dark Matter Distributions}, ApJ, 902, 98 (2020)

\bibitem{Gonzalez13}Gonzalez, A.H., Sivanandam,
S., Zabludoff, A.I. et al., \emph{Tracing Cosmic Evolution with Clusters of Galaxies} http://www.sexten-cfa.eu/en/conferences/2013/details/34-SestoClusters2013.html, id.123 (2013)



\bibitem{Huang16} Huang, Y., Lin, X.-W., Yuan, H.-B. et al., 
\emph{The Milky Way's rotation curve out to 100 kpc and its constraint on the Galactic mass distribution}, MNRAS, 463. 2623 (2016)

\bibitem{Jiao23} Jiao, Y., Hammer, F., Wang, H. et al., \emph{Detection of the Keplerian decline in the Milky Way rotation curve}, Astron. Astrophys. 678, A208 (2023)

\bibitem{Jesus18} Jesus, J.F., \emph{Exact solution for flat scale-invariant cosmology}
Rev. Mex. Astron. Astrophys. 55, 17, (2018)


 \bibitem{Karachantsev66} Karachantsev, I. D. 1966, Astrofizika, 2, 81, (In English: 1967, Astrophysics, 2, 39).
 
 
\bibitem{Kroupa02} Kroupa, P.,\emph{Thickening of galactic discs through clustered star formation}, MNRAS, 330, 707 (2002)


\bibitem{Kumamoto17} Kumamoto, J., Baba, J., Saitoh, T.R.,
\emph{Imprints of zero-age velocity dispersions and dynamical heating on the age-velocity dispersion relation}, Publ. Astron. Soc. Japan, 69, 31 (2017)


\bibitem{Lacey85} Lacey, C.G., Ostriker, J.P., \emph{Massive black holes in galactic halos ?},
ApJ, 299, 633 (1985)
 
 \bibitem{Landau60} Landau, L. \& Lifchitz, E.  Mecanique, Ed. Mir Moscou, 236 p. (1960)
 
 
\bibitem{Lang17} Lang, P., Forster Schreiber, N.M., 
Genzel, R., \emph{Falling Outer Rotation Curves of Star-forming Galaxies at 0 <z < 2.6 Probed with KMOS3D and SINS/zC-SINF},
ApJ, 840, 92L (2017)


 \bibitem{Leauthaud12} Leauthaud, A., 
George, M.R., Behroozi, P.S. et al., \emph{The Integrated Stellar Content of Dark Matter Halos}, ApJ, 746, 95 (2012)


\bibitem{Lelli17} Lelli, F., McGaugh, S.S., Schombert, J.M.,
 Pawlowski, M.S., \emph{One Law to Rule Them All: The Radial Acceleration Relation of Galaxies}, ApJ, 836, 152 (2017)
 
\bibitem{Lin12}Lin, Y.-T., Stanford, S.A.,
Eisenhardt,P.R.M. et al.,  \emph{ Baryon Content of Massive Galaxy Clusters at z = 0-0.6}, ApJ, 745, L3 (2012)







\bibitem{Maeder78} Maeder, A., \emph{Metrical connection in space-time, Newton's and Hubble's laws}, Astron. Astrophys. 65, 337 (1978)


\bibitem{Maeder17a} Maeder, A.,
\emph{An Alternative to the LambdaCDM Model: The Case of Scale Invariance}
Astrophys. J., 834, 194 (2017a)


\bibitem{Maeder17c} Maeder, A., \emph{
Dynamical Effects of the Scale Invariance of the Empty Space: The Fall of Dark Matter?}
Astrophys. J., 849, 158 (2017c)

\bibitem{Maeder19}
Maeder, A..\emph{ Evolution of the early Universe in the scale invariant theory},
arXiv:1902.10115 (2019)

\bibitem{Maeder23} Maeder, A., \emph{MOND as a peculiar case of the SIV theory},  MNRAS, 520, 1447  (2023)


\bibitem{MBouvier79} Maeder, A., Bouvier, P., \emph{ Scale invariance, metrical connection and the motions of astronomical bodies},
 Astron. Astrophys., 73, 82 (1979)


\bibitem{MaedGueor19} 
Maeder, A.; Gueorguiev, V.G., \emph{The growth of the density fluctuations in the scale-invariant vacuum theory},
Physics of the Dark Universe, 25, 100315 (2019)


\bibitem{MaedGueor20a}
Maeder, A.; Gueorguiev, V.G., 
\emph{The Scale-Invariant Vacuum (SIV) Theory: A Possible Origin of Dark Matter and Dark Energy}, 
Universe, vol. 6, 46 (2020a)

\bibitem{MaedGueor20b}  
Maeder, A.; Gueorguiev, V.G., \emph{Scale-invariant dynamics of galaxies, MOND, dark matter, and the dwarf spheroidal}, 
MNRAS, 492, 2698 (2020b)

\bibitem{MaedGueor21a}  
Maeder, A.; Gueorguiev, V.G., \emph{Scale invariance, horizons, and inflation},  MNRAS, 504, 4005 (2021a)




\bibitem{MaedGueor23}
Maeder, A.; Gueorguiev, V.G., \emph{Action Principle for Scale Invariance and Applications (Part I)}, Symmetry 2023, 15, 1966; arXiv 2310.16913 (2023)

\bibitem{Magnenat78} Magnenat, P., Martinet, L., Maeder, A., \emph{The age dependence of stellar velocity dispersion in a scale-covariant (theory of) gravitation}, Astron. Astrophys., 67, 51 (1978)

\bibitem{Mantz22} Mantz, A.B., Morris, R.G., Allen, S.V., \emph{Cosmological constraints from gas mass fractions of massive, relaxed galaxy clusters}, MNRAS, 510, 131 (2022)





\bibitem{Milgrom83} Milgrom, M., \emph{A modification of the newtonian dynamics : implications for galaxy systems}, Astrophys. J., 270, 365 (1983)


\bibitem{Milgrom09} Milgrom, M., \emph{The Mond Limit from Spacetime Scale Invariance}, Astrophys. J., 698, 1630 (2009)




\bibitem{Milgrom14} Milgrom, M., \emph{The MOND paradigm of modified dynamics}, Scholarpedia, 9, 31410 (2014a)













\bibitem{Nestor Shachar23} Nestor Shachar, A., Price, S.H., Forster Schreiber, N. M. et
al., \emph{RC100: Rotation Curves of 100 Massive Star-forming Galaxies at z = 0.6-2.5 Reveal 
Little Dark Matter on Galactic Scales}, ApJ 944, 78 (2023)

\bibitem{Nordstrom04} Nordstrom, B., Mayor, M., Andersen, J. et al., \emph{The Geneva-Copenhagen survey of the Solar neighbourhood. Ages, metallicities, and kinematic properties of 14 000 F and G dwarfs},
Astron. Astrophys. 418, 989 (2004)

\bibitem{Old18} Old, L., Wojtak, R., Pearce, F.R. et al., \emph{Galaxy Cluster Mass Reconstruction Project - III. The impact of dynamical substructure on cluster mass estimates},
MNRAS, 475, 852 (2018)




\bibitem{Planck20}
Planck Collaboration 2020, Aghanim, N., Akrami, Y. Ashdown, M., \emph{Planck 2018 results. VI. Cosmological parameters}, Astron. Astrophys. 641, A6 (2020)

\bibitem{Proctor15} Proctor, R.N.,  Mendes de Oliveira, C., Azanha, L., \emph{A derivation of masses and total luminosities of galaxy groups and clusters in the maxBCG catalogue},
 MNRAS, 449, 2345 (2015)

\bibitem{Raddi22} Raddi, R.,
Torres, S., Rebassa-Mansergas, A. et al., \emph{Kinematic properties of white dwarfs. Galactic orbital parameters and age-velocity dispersion relation},
Astron. Astrophys. 658, A22 R (2022)

\bibitem{Rubin78} Rubin, V.C., Ford, W.K. Jr., Thonnard, N., \emph{Radial velocities of spiral galaxies determined from 21-cm neutral hydrogen observations},  ApJ, 225, L107 (1978)


\bibitem{Seabroke07} Seabroke, G.M., Gilmore, 
  G., \emph{Revisiting the relations: Galactic thin disc age-velocity dispersion relation},  MNRAS, 380, 1348 (2007)
  

  
  \bibitem{Shan15} Shan, Y., McDonald, M., Courteau, S. , \emph{Revised Mass-to-light Ratios for Nearby Galaxy Groups and Clusters},  ApJ, 800, 122
  (2015)

  \bibitem{Sofue01} Sofue, Y., Rubin, V., \emph{Rotation Curves of Spiral Galaxies},
Annual Rev. Astron. Astrophys., 39, 137 (2001)
  
  \bibitem{Sohn17} Sohn, J., Geller , M.J., Zahid, H.J.
  et al., \emph{The Velocity Dispersion Function of Very Massive Galaxy Clusters: Abell 2029 and Coma}, ApJ Suppl., 229, 20  (2016)
  
  
  \bibitem{Spitzer51} Spitzer, L., Schwarzschild, M., \emph{The Possible Influence of Interstellar Clouds on Stellar Velocities}, ApJ, 114, 385 (1951)


 \bibitem{Toth92} Toth, G., Ostriker, J.P., \emph{Galactic Disks, Infall, and the Global Value of Omega}, ApJ, 389, 5 (1992)

\bibitem{Trimble87} Trimble, V., \emph{Existence and nature of dark matter in the universe}, Annual Rev. Astron. Astrophs., 25, 425 (1987)

\bibitem{Weinberg72} Weinberg, S.  \emph{Gravitation and Cosmology: Principles and applications of the General Theory
 of Relativity},  John Wiley \& Sons, Inc, New York, London, Sydney, Toronto
(1972)

 
 
\bibitem{Weyl23} {Weyl}, H., \emph{Raum, Zeit, Materie. Vorlesungen {\"u}ber allgemeine
Relativit{\"a}tstheorie}, Re-edited 1970 by Springer Verlag, Berlin, (1923)

\bibitem{Wicker23} Wicker, R., Douspis, M., Salvati, L. et al., \emph{Constraining the mass and redshift evolution of the hydrostatic mass bias using the gas mass fraction in galaxy clusters},  Astron. Astrophys.
674, A48 (2023)
  
 \bibitem{Wielen77} Wielen, R. 1977, \emph{The Diffusion of Stellar Orbits Derived from the Observed Age-Dependence of the Velocity Dispersion},
   Astron. Astrophys., 60, 263
  
 
 \bibitem{Zwicky33} Zwicky, F., \emph{Die Rotverschiebung von extragalaktischen Nebeln}, Helevetica Physica Acta,
 6, 10 (1933)











\end{thebibliography}
\end{document}